\documentclass[10pt]{iopart}
\expandafter\let\csname equation*\endcsname\relax
\expandafter\let\csname endequation*\endcsname\relax

\usepackage{amsmath}
\usepackage{graphicx}
\usepackage{txfonts}
\usepackage{color}
\usepackage{hyperref}

\newcommand{\rp}[1]{(\ref{#1})}

\newcommand{\abs}[1]{\left|{#1}\right|}

\newcommand{\av}[1]{\left\langle #1 \right\rangle}

\newcommand{\al}[1]{^{(#1)}}
\newcommand{\da}{^\dagger}

\newcommand{\pp}[1]{\left( #1 \right)}
\newcommand{\pq}[1]{\left[ #1 \right]}
\newcommand{\pg}[1]{\left\{ #1 \right\}}

\newcommand{\lpg}[1]{\left\{ #1 \right.}

\newcommand{\rpg}[1]{\left. #1 \right\}}

\newcommand{\ii}{{\rm i}}
\newcommand{\dd}{{\rm d}}

\newcommand{\nn}{{\nonumber}}

\newcommand{\CC}{{\cal C}}

\newcommand{\GG}{{\cal G}}

\newcommand{\QQ}{{\cal Q}}

\newcommand{\TT}{{\cal T}}

\usepackage[normalem]{ulem}

\begin{document}

\title{
An optomechanical heat engine with feedback-controlled in-loop light
}

\author{Najmeh Etehadi Abari$^{1,2}$, Giulia Vittoria De Angelis$^{1}$, Stefano Zippilli$^{1}$, David Vitali$^{1,3,4}$}
\address{
$^{1}$ School of Science and Technology, Physics Division, University of Camerino, I-62032 Camerino (MC), Italy\\
$^{2}$ Department of Physics, Faculty of Science, University of Isfahan, Hezar Jerib, 81746-73441, Isfahan, Iran\\
$^{3}$ INFN, Sezione di Perugia, I-06123 Perugia, Italy\\
$^{4}$ CNR-INO, Largo Enrico Fermi 6, I-50125 Firenze, Italy}
\vspace{10pt}
\begin{indented}
\item[]\today
\end{indented}

\begin{abstract}
The dissipative properties of an optical cavity can be effectively controlled by placing it in a feedback loop where the light at the cavity output is detected and the corresponding signal is used to modulate the amplitude of a laser field which drives the cavity itself.
Here we show that this effect can be exploited to improve the performance of an optomechanical heat engine which makes use of polariton excitations as working fluid. In particular we demonstrate
that, by employing a positive feedback close to the instability threshold, it is possible
to operate this engine also under parameters regimes which are not usable without feedback, and which may significantly ease the practical implementation of this device.
\end{abstract}

%
%
%
%
%


\section{Introduction}

Heat engines convert thermal energy into work.
A quantum heat engine uses a quantum system as working fluid.
The practical realization of these devices is interesting as platforms for the experimental investigation of the thermodynamics of the quantum world and of non-equilibrium systems~\cite{vinjanampathy2016,alicki2018}.

Optomechanics~\cite{bowen2015} describes systems, which range from the nanoscale to macroscopic sizes, where the interaction between light and mechanical objects is exploited for enhanced metrology \cite{Li2018}, and to explore the limits of quantum physics \cite{aspelmeyer2014,Bawaj2015}.
Thermal machines based on optomechanical systems have been proposed and analyzed in different configurations~\cite{zhang2014a,zhang2014,dong2015,dong2015a,dechant2015a,mari2015,gelbwaser-klimovsky2015a,bathaee2016,zhang2017}.
A specific example~\cite{zhang2014a,zhang2014,dong2015} makes use of hybridized polariton excitations as working fluid. This engine works in the strong optomechanical coupling regime where the normal modes of the system are superpositions of optical and mechanical excitations.
This regime is in general not easily achievable and in some cases is inhibited by detrimental non-linear processes, such as optical bistability or thermorefractive effects, which hamper the ability to carefully control the coupled dynamics of the systems.
It has been shown~\cite{rossi2018} that feedback-controlled light~\cite{rossi2017,kralj2017,rossi2018,zippilli2018} can be employed to significantly ease the onset of strong coupling in an optomechanical system.
This suggests~\cite{deangelis2018} that the feedback analyzed in~\cite{zippilli2018} can be used to enhance the efficiency of the quantum heat engine proposed in\cite{zhang2014a}.

In this article we analyze the effect of feedback-controlled light on the performance of the polariton-based optomechanical heat engine discussed in~\cite{zhang2014a,zhang2014,dong2015}. We show that, with the aid of feedback, this engine can operate efficiently also when the system is not in the strong coupling regime and for parameters for which, in the absence of feedback, the engine is not functional.

The article is organized as follows. In section~\ref{model} we introduce the model of the optomechanical system driven by a feedback-controlled  pump field. In section~\ref{heat_engine}  we review the functioning of the quantum heat engine introduced in~\cite{zhang2014a,zhang2014,dong2015}. Then, in section~\ref{feedbak_engine}
we discuss the effect of feedback on the performance of this device. In section~\ref{upper} we present a variant of the engine which exploits the upper polariton mode as working fluid. Finally, section~\ref{conclusion} is for the conclusions.

\section{The model}\label{model}

\begin{figure}
\centering
\includegraphics[width=12cm]{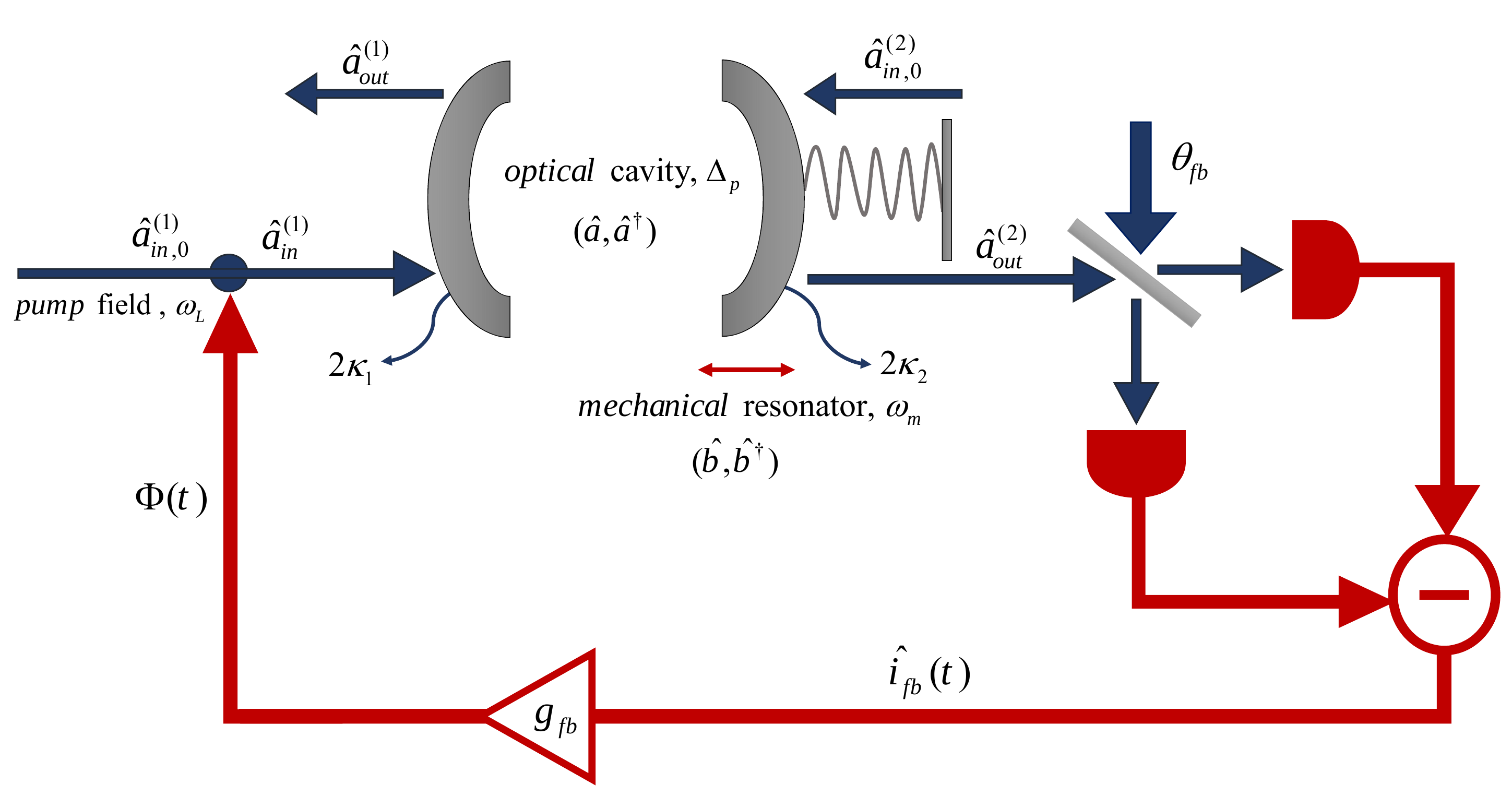}
\caption{The feedback loop: a quadrature of the field, transmitted through a Fabry-P\'erot cavity with a movable end mirror, is detected via homodyne detection at phase $\theta_{fb}$, and the corresponding photocurrent is used to modulate the input field \cite{zippilli2018}.}
\label{Fig1}
\end{figure}

In this work we consider an optomechanical device similar to the one discussed in Refs.~\cite{zippilli2018}, composed of an optical cavity with a moving end mirror placed within a feedback loop where the light transmitted through the cavity is detected and the corresponding signal is used to modulate the amplitude of the laser field which drives the system, see Fig.~\ref{Fig1}.
In details, one resonant mode of the optical cavity at frequency $\omega_c$ and with decay rate $\kappa_c$, is coupled
to a vibrational mode of the mirror, at frequency $\omega_m$, which dissipates its energy at rate $\gamma\ll\omega_m$.
The laser is at frequency $\omega_L$ and is detuned by $\Delta_p=\omega_L-\omega_c$ form the cavity resonance.
We describe the system in terms of the standard linearized model for the fluctuations of the optical and mechanical variables about the corresponding average values~\cite{bowen2015} (this also implies that the cavity frequency includes the shift due to the optomechanical interaction). 
Specifically, assuming that the feedback does not affect the average laser intensity (this
can be realized using a high-pass feedback response function which cuts the low-frequency components of the photocurrent~\cite{zippilli2018}), the annihilation and creation operators for optical and mechanical excitations fulfill the quantum Langevin equations
\begin{eqnarray}\label{QLE0}
\dot{\hat{a}} &=&  - (\kappa _c  - i \Delta _p )\hat a - i G (\hat b + \hat b^{\dagger} ) + \sqrt {2\kappa _c } \hat a_{in},  \\
\dot{\hat{b}} &=&  - (\gamma  + i \omega_m )\hat b - i G (\hat a + \hat a^{\dagger}) + \sqrt {2\gamma } \hat b_{in},
\label{dotb}
\end{eqnarray}
where $G$ is the linearized coupling strength,  $\hat b_{in} (t)$ is the noise operator for the mechanical resonator which describes thermal noise with $n_{th}$ thermal excitations according to the correlation function $\av{\hat b_{in}(t)\ \hat b_{in}{}\da(t') }=(1+n_{th})\delta (t - t')$, $\hat a_{in} (t)$ is the input noise operator for the cavity field which can be decomposed in terms of the noise operators $\hat a_{in}^{(1)} (t)$ and $\hat a_{in}^{(2)} (t)$ associated with the left and the right mirror respectively, as
\begin{eqnarray}
\hat a_{in} (t) &=& \frac{{\sqrt {2\kappa _1 } \: \hat a_{in}^{(1)} (t) + \sqrt {2\kappa _2 } \: \hat a_{in}^{(2)} (t)}}{{\sqrt {2\kappa _c } }},
\end{eqnarray}
with $\kappa_1$ and $\kappa_2$ the corresponding decay rates, such that $\kappa _c = \kappa_1+\kappa_2$. In turn, the noise operator $\hat a_{in}^{(1)} (t)$ can be decomposed as the sum of the operator without feedback plus an additional term $\hat \Phi(t)$
due to the feedback
$\hat a_{in}^{(1)} (t) = \hat a_{in,0}^{(1)} (t) + \hat \Phi (t)$.
The input noise operators $\hat a_{in,0}^{(1)} (t)$ and $\hat a_{in}^{(2)} (t)$ describe vacuum fluctuations and are characterised by the correlation functions
$\av{\hat a_{in,0}^{(1)}(t)\ \hat a_{in,0}^{(1)}{}\da(t') }=\delta (t - t')$ and
$\av{\hat a_{in}^{(2)}(t)\ \hat a_{in}^{(2)}{}\da(t') }=\delta (t - t')$.
The feedback term $\hat \Phi(t)$ depends on the feedback photocurrent~\cite{zippilli2018}. In particular, if the feedback gain $\bar g_{fb}$ is constant over a sufficiently large band of frequencies around the mechanical resonance, it can be approximated as $\hat \Phi (t) = \bar g_{fb} \: \hat i_{fb} (t - \tau _{fb})$, such that it is proportional to the photocurrent $\hat i_{fb}(t)$ at an earlier time determined by the feedback delay time $\tau_{fb}$~\cite{zippilli2018,deangelis2018}, so that
\begin{eqnarray}\label{ain1}
\hat a_{in}^{(1)} (t) &=& \hat a_{in,0}^{(1)} (t) + \bar g_{fb} \: \hat i_{fb} (t - \tau _{fb})\ .
\end{eqnarray}
The photocurrent resulting from the homodyne detection of the field at the output of the second mirror is expressed as~\cite{zippilli2018}
\begin{eqnarray}\label{i}
\hat i_{fb} (t) &=& \sqrt {\eta _d } \: \hat X_{out,fb}^{(\theta _{fb} )} (t) + \sqrt {1 - \eta _d } \: \hat X_\nu  (t),
\end{eqnarray}
where $ \theta_{fb} $ is the phase of the local oscillator, $\eta_d $ is the detection efficiency,
$ \hat X_\nu  (t)
$ is an operator representing additional noise due to the inefficient detection, which satisfies the relation $ \langle {\hat X_\nu  (t)\hat X_\nu  (t')} \rangle  = \delta (t - t') $, and
$\hat X_{out,fb}^{(\theta _{fb} )} (t) = e^{ - i\theta _{fb} } \: \hat a_{out}^{(2)} (t) + e^{  i\theta _{fb} } \: \hat a_{out}^{(2)}{}^{\dagger} (t) $ is the detected field quadrature at phase $\theta_{fb}$, with corresponding annihilation operator determined by the standard input-output relation~\cite{gardiner2004}
\begin{eqnarray}\label{in-out}
 \hat a_{out}^{(2)} (t) &=& \sqrt {2\kappa _2 } \: \hat a(t) - \hat a_{in}^{(2)} (t) .
\end{eqnarray}
According to Eqs.~\rp{ain1} and \rp{i} this operator is calculated at the delayed time $\hat a_{out}\al{2}(t-\tau_{fb})$ and,
in the regime of large detuning with respect to the optomechanical coupling constant and cavity decay rate, i.e., $| {\Delta _p } | \gg G,\kappa_c$, it is convenient to rewrite it
as a product of two terms (a slowly varying one and fast oscillating one) as
$\hat a_{out} (t - \tau _{fb} ) = \hat {\bar{ a}}_{out} (t - \tau _{fb} )\: e^{ - i\Delta _p (t - \tau _{fb} )}$.
Whenever the delay time is much shorter than both the characteristic time of the interaction $1/G$ and the decay time of the cavity $1/ 2 \kappa_c $, i.e., $\tau_{fb} < 1/G  , 1/2 \kappa_c$, we can ignore the delay time dependence of the slow part $\hat{\bar a}_{out} (t - \tau _{fb} )$ and then rewrite the output operator as
\begin{eqnarray}
\hat a_{out} (t - \tau _{fb} ) &\simeq& \hat {\bar{ a}}_{out} (t) \: e^{ - i\Delta _p t} \: e^{i\Delta _p \tau _{fb} }  = \hat a_{out} (t) \: e^{i\Delta _p \tau _{fb} }.
\end{eqnarray}
In this situation the delay-time dependence of the photocurrent in Eq.~\rp{ain1} can be approximated as a phase factor
such that
\begin{eqnarray}\label{i2}
\hat i_{fb} (t - \tau _{fb} ) &=& \sqrt {\eta _d } \: \left( e^{ - i\phi } \: \hat a_{out}^{(2)} (t) + e^{i\phi } \: \hat a_{out}^{(2)}{}^{\dagger} (t) \right) + \sqrt {1 - \eta _d } \: \hat X_\nu  (t),
\end{eqnarray}
where we have introduced the global phase $\phi \equiv \theta_{fb}-\Delta_p \tau_{fb}$.

By using Eqs.(\ref{in-out}), (\ref{ain1})and (\ref{i2}) and assuming $\phi=0$ (this can be achieved by properly adjusting the value of $\theta_{fb}$ depending on the value of detuning), we can rewrite the equation for the cavity operator~\rp{QLE0} as
\begin{eqnarray}\label{dota}
\dot{\hat{a}}(t) &=&  - (\kappa _{fb}  - i\Delta _{p} )\: \hat a(t) + (\kappa_c-\kappa_{fb})  \:  \hat a^{\dagger}(t)  - iG \: \pq{\hat b(t) + \hat b^{\dagger}(t)} +\sqrt{2 \kappa_{fb}} \: \hat a_{in,fb} (t),
\end{eqnarray}
where we have introduced the feedback-modified cavity decay rate
\begin{eqnarray}\label{eff0}
\kappa _{fb}  &=& \kappa _c  - 2 \bar g_{fb} \sqrt {\eta _d \kappa _1 \kappa _2 }
\end{eqnarray}
and the corresponding noise operator
\begin{eqnarray}
\hat a_{in,fb} (t) &=& \frac{1}{{\sqrt {2\kappa _{fb} } }} \lpg{
\sqrt {2\kappa _1 } \: \hat a_{in,0}^{(1)} (t) + \sqrt {2\kappa _2 } \: \hat a_{in}^{(2)} (t) - \bar g_{fb} \sqrt {2\eta _d \kappa _1 } \: \left[ e^{ - i\phi } \hat a_{in}^{(2)} (t) + e^{i\phi } \hat a_{in}^{(2)}{}^{\dagger} (t) \right]
}\nn\\&&\rpg{
+ \bar g_{fb} \sqrt {2(1 - \eta _d )\kappa _1 } \ \hat X_\nu  (t),
}\ ,
\end{eqnarray}
which describes additional effective thermal noise characterised by the correlation relations
\begin{eqnarray}
\av{\hat a_{in,fb}^{\dagger}(t) \: \hat a_{in,fb} (t')}  &=& n_{opt,fb} \: \delta (t - t')\ , \\
\av{\hat a_{in,fb} (t) \: \hat a_{in,fb}(t')} &=& 0\ ,
\end{eqnarray}
with the feedback-mediated number of thermal excitations  defined as
\begin{eqnarray}\label{nop3}
n_{opt,fb}  &=& \frac{{(\kappa _c  - \kappa _{fb} )^2 }}{{\eta _d \kappa _c \kappa _{fb} }}.
\end{eqnarray}
This shows that, the feedback-controlled system behaves as an effective optomechanical system with modified cavity decay rate $\kappa_{fb}$, under the effect of additional noise with a finite number of thermal excitations $n_{opt,fb}$ and of an additional parametric driving term with strength $\kappa_c-\kappa_{fb}$ [see Eq.~\rp{dota}]. The values of both $\kappa_{fb}$ and $n_{opt,fb}$ are controllable via the feedback gain $\bar g_{fb}$ according to the relations~\rp{eff0} and \rp{nop3}. This allows one to operate the same system under different parameter regimes~\cite{rossi2017,kralj2017,rossi2018,zippilli2018}.
In particular when the feedback is operated close to its instability threshold, namely when the effective cavity decay rate becomes very small ($\kappa_{fb}\to 0$),
also a weakly coupled system may exhibits the typical features of strongly coupled systems such as normal mode splitting~\cite{rossi2018}.

\section{The polariton-based optomechanical heat engine}\label{heat_engine}

\begin{figure} \centering
\includegraphics[width=8cm]{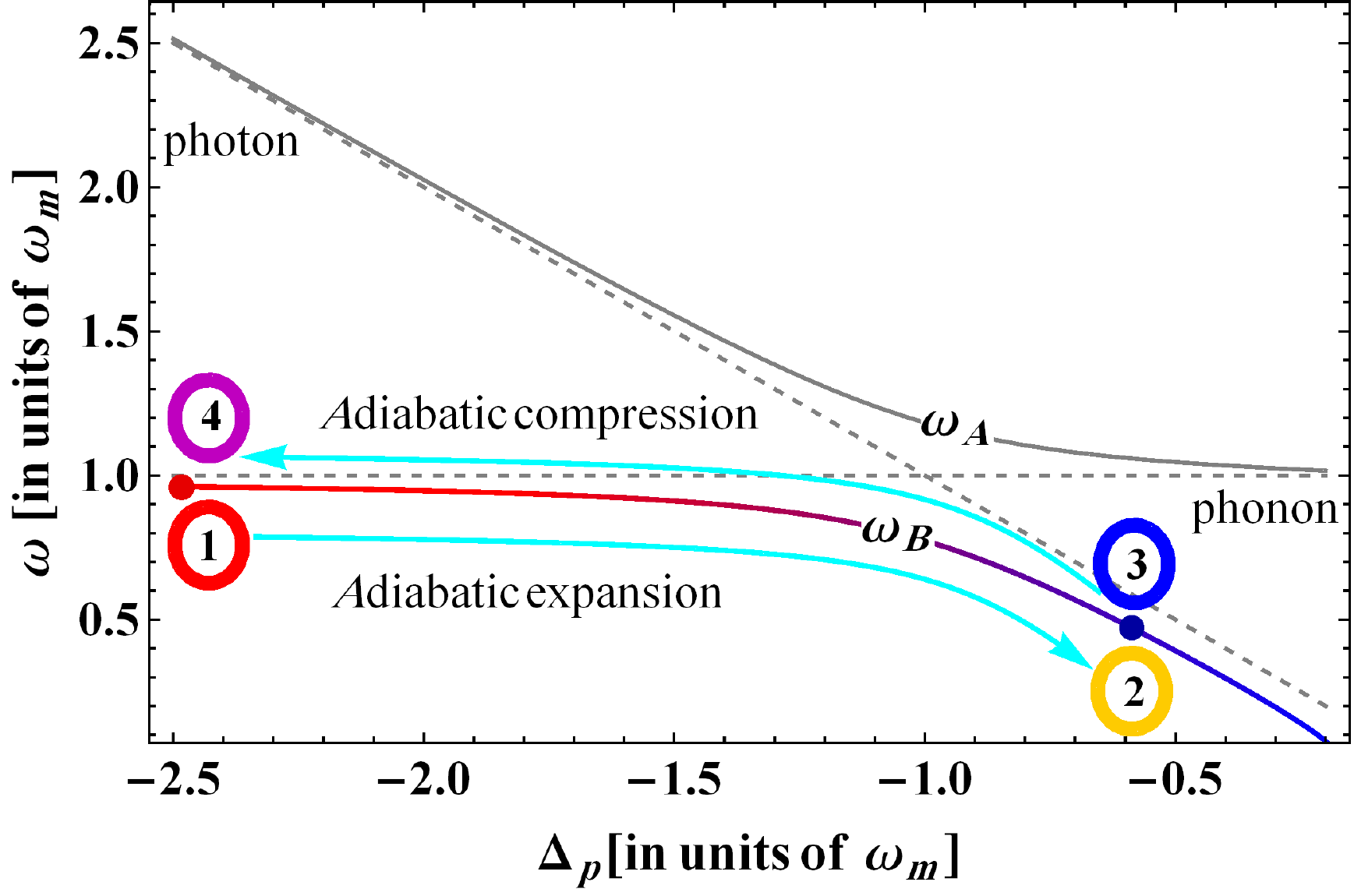}
\caption{
Frequency of the two polaritons (upper $\omega_A$ and lower $\omega_B$) of the optomechanical system as a function of the cavity detuning $\Delta_p$ in the red-detuned case $\Delta_p < 0$ (the optomechanical coupling strength is $G= 0.05\,\omega_m$ ). The dashed curves correspond to the frequencies of the  non-interacting modes. In the plot we have indicated the position of the four nodes of the Otto cycle operated on the lower polariton. The strokes from node $1$ to $2$ and from $3$ to $4$ correspond to the adiabatic processes. The strokes from node $2$ to $3$ and from $4$ to $1$ take place at constant detuning and correspond to the isochoric processes.
}
\label{Fig2}
\end{figure}

References~\cite{zhang2014a,zhang2014,dong2015} describe a quantum heat engine which makes use of polariton excitations in an optomechanical system (without feedback) as working fluid. This device requires strong optomechanical coupling and resolved sideband regime $\omega_m,G\gg\kappa_c$ for its functioning.
And works at red detuning, where the laser frequency is lower than the cavity frequency so that the optical and mechanical mode can exchange coherently their excitations.
Only in this regime the hybridised polariton excitations (the excitations of the normal modes of the system) become relevant.
The engine focuses on the lower polariton mode and realises an Otto cycle formed by two adiabatic and two isochoric processes.
Specifically it works as follows (see Fig.~\ref{Fig2}).
The lower polariton frequency plays the role of the volume of the working fluid (similar to other single oscillator engines~\cite{vinjanampathy2016,alicki2018}) and it can be controlled via the laser detuning. This is the central tool used to operate the engine through the four strokes of the cycle.
At large detuning the lower polariton is phonon-like and it
is in thermal contact
with the hot mechanical thermal bath (see Fig.~\ref{Fig2}).
A fast change of the detuning brings the laser closer to the cavity resonance, passing through the red mechanical sideband, until the polariton becomes photon-like and comes into contact with the cold (zero temperature) optical bath. This variation of the detuning realizes the first adiabatic process from 1 to 2 (see Fig.~\ref{Fig2}). Hence it has to be sufficiently fast in order to avoid dissipation.
However, at the same time, it has to be sufficiently slow in order to avoid non-adiabatic transitions to the upper polariton mode. This means that the duration $\tau_1$ of this process must fulfil the conditions
$1/G\ll\tau_1\ll 1/\kappa_c$
(note that the linearized optomechanical coupling is assumed constant during the cycle;
this can be achieved by properly controlling the pump intensity during the adiabatic processes~\cite{zhang2014a,zhang2014,dong2015}). After the adiabatic process, the detuning is kept fixed at the value closest to the cavity resonance for a time $\tau_2\gg 1/\kappa_c$ until the lower (photon-like) polariton thermalizes with the optical reservoir realising the first isochoric process.
This process has to be sufficiently short ($\tau_2\ll 1/\gamma$) in order to avoid mechanical dissipation of the upper (phonon-like) polariton which should not contribute to the variation of the system energy during the cycle.
The second adiabatic process is realized by sweeping back the detuning to the initial value over a time $\tau_3=\tau_1$ so to guarantee the adiabaticity of the process.
Now, the lower polariton is again phonon-like and in the second isochoric process it thermalizes with the thermal mechanical bath, over a time $\tau_4\gg 1/\gamma$. The upper polariton, instead, does not change significantly its number of excitations during the full cycle.

\section{The feedback-enabled heat engine}\label{feedbak_engine}

A critical requirement in this device is the ability to realise the adiabatic processes which needs a sufficiently large difference between $G$ and $\kappa_c$. This is the regime of strong coupling that, although reached in a few systems~\cite{Groblacher2009a,Teufel2011}, is not straightforward to achieve, and in certain cases it is inhibited by the onset of detrimental non-linear effects~\cite{rossi2018}.
As discussed in Ref.~\cite{rossi2018}, the feedback that we have described above seems particularly fit for this purpose, and can be exploited
to ease the realisation of this engine.
In particular, on the one hand the feedback loop enables one to reduce the cavity bandwidth and to bring a system into the strong coupling regime even if naturally it is weakly coupled; on the other hand it adds extra noise to the cavity, corresponding to a finite number of thermal photonic excitations $n_{opt,fb}$.
In order to realize the heat engine that works on the lower polariton, the cycle should work between a cold photonic reservoir and a hot phononic reservior. This means that the Otto cycle that we have discussed can be effective as long as $n_{opt,fb} < n_{th}$, which implies [see Eq.~\rp{nop3}] that $\kappa_{fb}$ cannot be too small.
Furthermore, a difference between the model of~\cite{zhang2014a,zhang2014} and the feedback-controlled system introduced in Sec.~\ref{model} is the additional parametric driving in the latter [see Eq.~\rp{dota}]. However when the system is in the resolved sideband regime its effect is very small.
Hence, neglecting the parametric term we can perform an analysis similar to the one discussed above  also in the case of feedback.
And we can state that, when one utilises feedback,
the engine can work efficiently when the duration of the four strokes fulfill the following set of relations
\begin{eqnarray}\label{hi}
&&\frac{1}{G}\ll\tau_1,\tau_3 \ll \frac{1}{\kappa_{fb}}\ll\tau_2\ll\frac{1}{\gamma}\ll\tau_4\ ,\\
\end{eqnarray}
and
\begin{eqnarray}
&&\hspace{1.3cm}n_{opt,fb} < n_{th}\ .
\end{eqnarray}

If the cycle operates optimally with ideal adiabatic passages, then it realises a perfect Otto cycle where,
in the adiabatic processes, the system exchanges energy with the environment in terms of work without transferring
heat, instead, in the isochoric processes the system exchanges only heat and thermalizes with the environment (see~\ref{heat_work} for a definition of heat and work that applies to this system).
In this case the heat and the work in each stroke is given by the difference between the system's energy at the beginning and at the end of each stroke $\Delta E_{i\to j}=E_j-E_i$ (for $i,j=1,2,3,4$), where, denoting with the label $A$ the upper polariton and with $B$ the lower one,
the system energy is given by $E=\hbar\pp{\omega_A N_A+\omega_B N_B}$, with $\omega_x$ and $N_x$ (for $x\in\pg{A,B}$) the frequency and the number of polariton excitations respectively.
The polariton $A$ is initially photon-like and its frequency is given by the initial cavity detuning $\omega_A\sim\abs{\Delta_{i}}$, with corresponding number of excitations $N_A\sim n_{opt,fb}$.
The polariton $B$, instead, is initially phonon like with frequency $\omega_B\sim\omega_m$ and $N_B\sim n_{th}$ excitations.
At the end of the first adiabatic process $A$ becomes phonon-like, at frequency $\omega_A\sim\omega_m$, and $B$ photon-like with a frequency close to the corresponding cavity detuning $\omega_B\sim\abs{\Delta_{f}}$.
Then, in the isochoric process the polariton $B$ thermalizes with the feedback-mediated optical bath, while $A$ should remain with its initial number of excitations.
Then, in the second adiabatic process the polariton frequencies return to their initial values, and finally in the second isochoric process, polariton $B$ returns to its initial value of excitations (note that during an ideal cycle the number of excitations of the polariton $A$ should remain constant).
Hence ideally the changes of energy (and the corresponding heat $Q$ and work $W$) in the four strokes, are
\begin{eqnarray}
W_{1\to2}=\Delta E_{1\to 2}&\sim&\hbar\pp{\abs{\Delta_{f}} n_{th}+\omega_m\ n_{opt,fb}}-\hbar\pp{\omega_m\ n_{th}+\abs{\Delta_{i}}\ n_{opt,fb}}<0,
\nn\\
Q_{2\to3}=\Delta E_{2\to 3}&\sim&\hbar\abs{\Delta_{f}} n_{opt,fb}-\hbar\abs{\Delta_{f}} n_{th}<0,
\nn\\
W_{3\to4}=\Delta E_{3\to 4}&\sim& \hbar\pp{\omega_m\ n_{opt,fb}+\abs{\Delta_{i}}\ n_{opt,fb}}-\hbar\pp{\abs{\Delta_{f}}\ n_{opt,fb}+\omega_m\ n_{opt,fb}}>0,
\nn\\
Q_{4\to1}=\Delta E_{4\to 1}&\sim&\hbar\ \omega_m\ n_{th}-\hbar\ \omega_m\ n_{opt,fb}>0.
\end{eqnarray}
The negative work in the first stroke indicates that the work is performed by the system, while the positive heat in the fourth stroke indicates that the heat is absorbed by the system.
The efficiency of the cycle is given by the ratio $\eta=-W_{tot}/Q_{abs}$ between the total work $W_{tot}=W_{1\to2}+W_{3\to4}$ and the absorbed heat $Q_{abs}=Q_{4\to1}$, that is,
\begin{eqnarray}\label{eta}
\eta=\frac{-\pp{W_{1\to2}+W_{3\to4}}}{Q_{4\to1}}\sim 1-\frac{\abs{\Delta_{f}}}{\omega_m}\ .
\end{eqnarray}
Note that during the adiabatic processes part of the work is also done by and on the polariton $A$ (respectively in the first and second adiabatic process). However, since the number of excitations does not change the net contribution to the total work due to polariton $A$ is zero.

\subsection{Results}\label{results}

\begin{figure} \centering
\centering
\includegraphics[width=11.5cm]{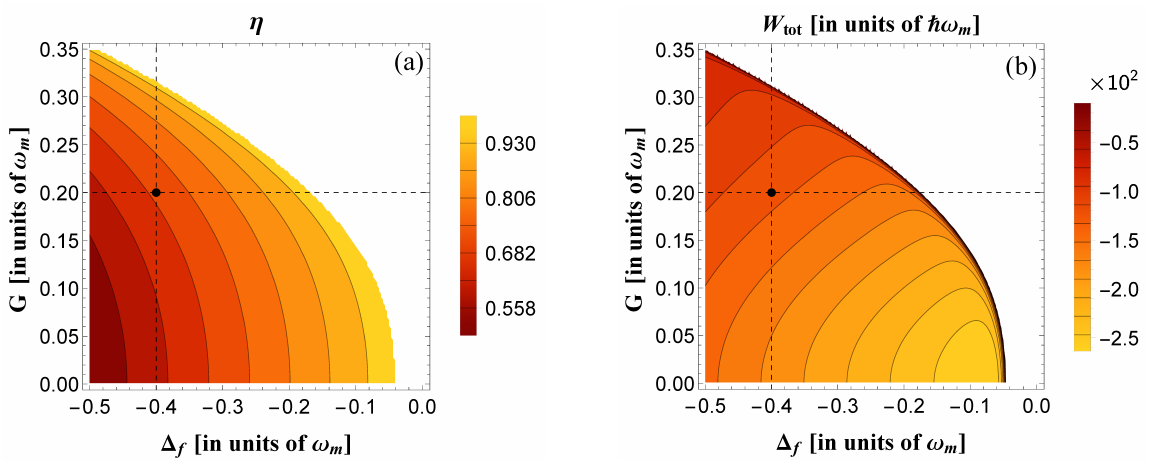}
\caption{
(a) Thermal efficiency $\eta$ and (b) total work $W_{tot}$ done by the engine
(operated on the lower polariton) as a function of the smallest detuning $\Delta_{f}$ and the optomechanical coupling strength $G$, evaluated in terms of the steady-state energy corresponding to the lower polariton mode at each node of cycle, assuming perfect adiabatic processes and constant excitations of the upper polariton mode. The white areas indicate the parameter regime at which the system is unstable (see~\ref{heat_work}).
The dots indicate the parameters used for the results in Figs.~\ref{Fig4} and \ref{Fig5}.
The other parameters are $\Delta_{i} = -3\omega_m$, $2 \kappa_c = 0.1 \omega_m$, $2 \gamma = 10^{-4} \omega_m$, and $n_{th} = 300$. The feedback is set in order to achieve the effective cavity decay rate $ 2 \kappa_{fb}=0.015\omega_m$, corresponding to an effective number of thermal photons $n_{opt,fb} \approx 8$ (see Eq. \ref{nop3}).
}
\label{Fig3}
\end{figure}

A more accurate estimate of the efficiency and of the work performed by this engine can be computed
by focusing on the steady state properties of the lower polariton $B$ alone at the end of each stroke, but still assuming perfect adiabatic processes and constant population of the upper polariton throughout the whole cycle.
This allows to better estimate the expected performance of the engine by taking into account also the effects of the optomechanical interaction at large and small detuning.
Specifically, Eqs.~\rp{dotb} and \rp{dota} can be used to evaluate the steady state correlation matrix $\CC_{ss}$ of the system (see~\ref{matrices} for details); moreover, the populations and the frequencies of the polariton mode $B$ can be estimated by transforming the correlation matrix to the polariton bases, which determines the normal modes of the system Hamiltonian as discussed in~\ref{polariton}.
 This allows to estimate the energy associated to the polariton $B$ at the beginning and end of each stroke ($E_{B,j}= \hbar \omega_{B,j}\,N_{B,j} $ for $j=1,2,3,4$) and to estimate the corresponding work and heat [such as $W_{tot}\sim-\pp{E_{B,2}-E_{B,1}+E_{B,4}-E_{B,3}}$ and $Q_{abs}\sim E_{B,1}-E_{B,4}$].
The corresponding results are reported in Fig.~\ref{Fig3}.
They show that the optimal performance
of the engine are achieved at small $\Delta_p$ and $G$~\cite{zhang2014a,zhang2014}.
In this regime however our estimate are likely to be inaccurate.
In fact, on the one hand, at vanishing $G$ the time for the adiabatic passage needs to be extremely long (longer then the dissipation time), and on the other hand at small $\Delta_p$ the effect of the parametric term can become important [in fact at very small $\Delta_f$ the system is unstable (see~\ref{heat_work}) as indicated by the white areas in Fig.~\ref{Fig3}]. In order to address this issue more rigorously we have analyzed the full dynamics of the system.

\begin{figure} \centering
\centering
\includegraphics[width=5cm]{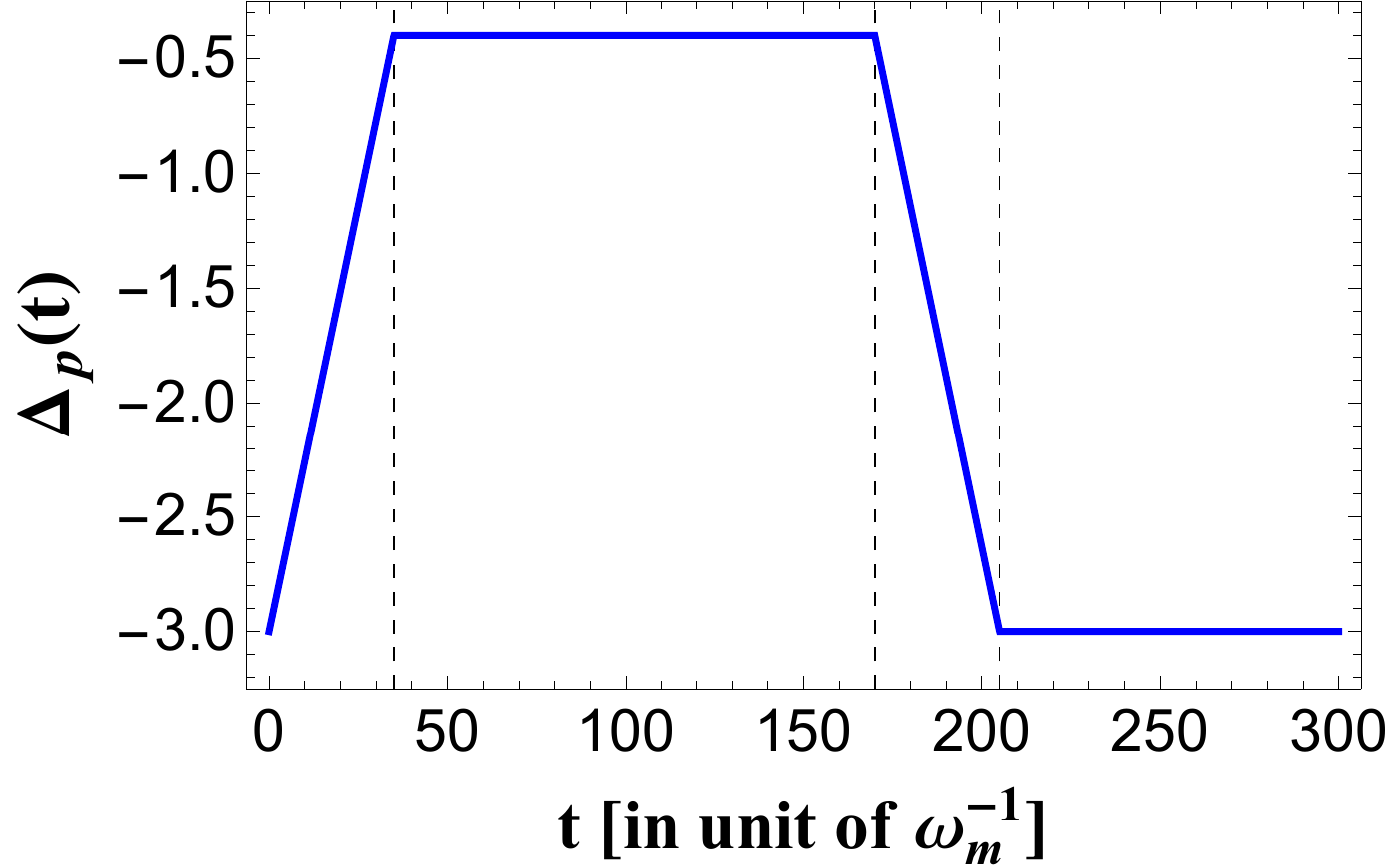}
\caption{
Time evolution of the detuning $\Delta_p$. The vertical dashed lines indicate the end and the beginning of each stroke. The last stroke 
is not shown completely because it is very slow due to the small mechanical damping rate $\gamma$. During the last stroke the $\Delta_p$ remains constant.
}
\label{Fig3-0}
\end{figure}

An in-depth study of the efficiency of the engine is achieved by solving the quantum Langevin equations Eqs~\rp{dotb} and \rp{dota},
and computing the time evolution of the energy exchanged between the system and the environment in terms of heat and work as discussed in~\ref{heat_work}. Specifically, these quantities can be expressed in terms of the correlation functions of the system operators, the dynamics of which can be computed by standard techniques (see~\ref{matrices}).
Hereafter we report and discuss the result of this numerical analysis when the cavity detuning $\Delta_p$ is changed in time, in order to realise the Otto cycle, according to the relation
\begin{eqnarray}\label{Delta}
\Delta _p (t) &=& \left\{ \begin{array}{ll}
 \frac{{\Delta _f  - \Delta _i }}{{t_1  - t_0 }}(t - t_0 ) + \Delta _i \ ,	& {\rm   t}_{\rm 0}  \le t < {\rm t}_{\rm 1}  \\
 \Delta _f \ ,									& {\rm   t}_{\rm 1}  \le t < {\rm t}_{\rm 2}  \\
 \frac{{\Delta _i  - \Delta _f }}{{t_3  - t_2 }} (t - t_2 ) + \Delta _f \ ,	& {\rm t}_{\rm 2}  \le t < {\rm t}_{\rm 2}  \\
 \Delta _i \ ,									& {\rm  t}_{\rm 3}  \le t \le {\rm t}_{\rm 4} . \\
 \end{array} \right.
\end{eqnarray}
In details (see Fig.\ref{Fig3-0}), in the first stroke the detuning is changed linearly from the initial value $\Delta_i$ to $\Delta_f$. Then it is kept constant at the value $\Delta_f$. In the third stroke it changes linearly back to the initial value. And finally, in the last stroke, it remains constant at the value $\Delta_i$. The duration of each stroke is $\tau_j=t_1-t_{j-1}$, for $j=1,2,3,4$.

\begin{figure} \centering
\includegraphics[width=12.5cm]{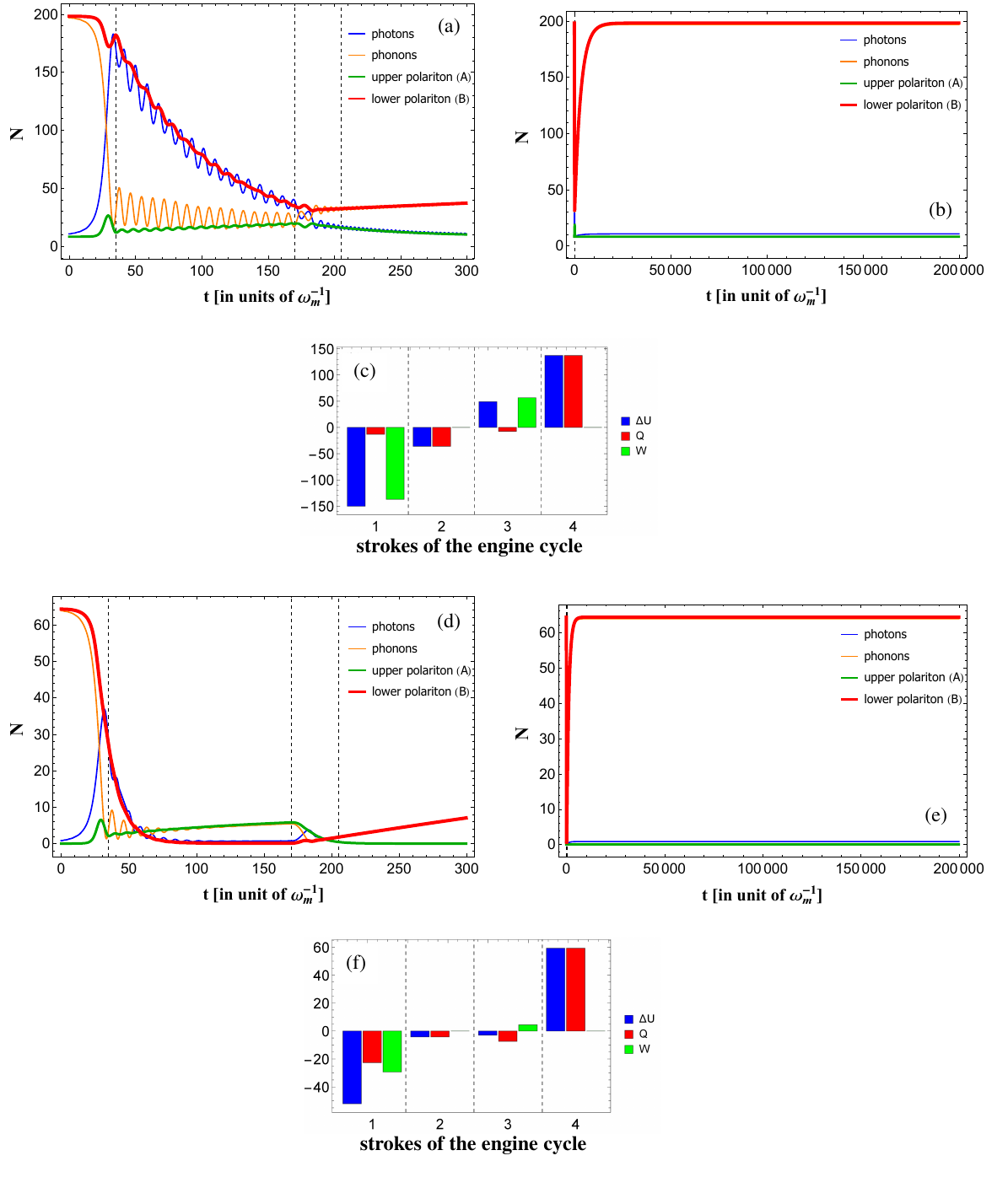}
\caption{
(a), (b), (d), (f) Time evolution of the populations of the polariton modes, $N_B$ (red line) and $N_A$ (green line), and of the photonic ($N_a$, blue line) and phononic modes ($N_b$, orange line) during a loop of the engine cycle (with initial state given by the system steady state at the initial detuning) with (a), (b) and without (d), (e) feedback. (a) and (d) show the dynamics in the first three strokes.
(b) and (d) show a longer time scale that highlight the slow thermalization in the fourth stroke. Corresponding energy change (blue), heat exchanged (red) and work performed (green) during each stroke of the cycle. The duration of each stroke is $\tau_1=\tau_3 = 35 \omega_m^{-1}$,  $\tau_{2} = 135 \omega_m^{-1}$ and  $\tau_{4} = 20/\gamma$. 
The other parameters are as in Fig.~\ref{Fig3}.
}
\label{Fig4}
\end{figure}

\begin{figure}[t] \centering
\includegraphics[width=15.5cm]{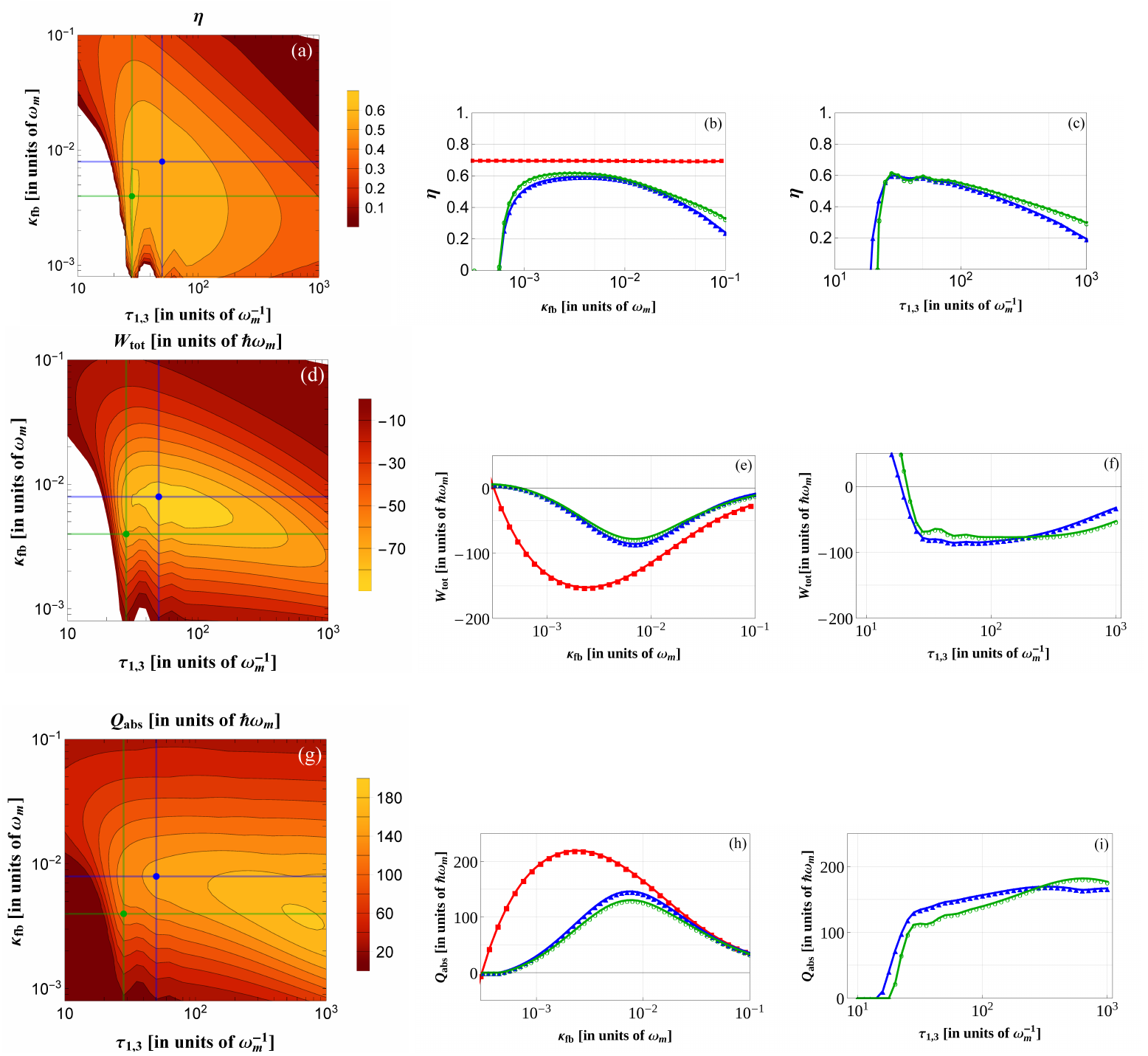}
\caption{
(a),(b),(c): Efficiency of the quantum engine; (d),(e),(f): total work; (g),(h),(i): absorbed heat during the cycle, versus $\kappa_{fb}$ and $\tau_1\ (=\tau_3)$, evaluated by computing the system time evolution as discussed in~\ref{heat_work}.
The plots in the second column report the values of efficiency (b), work (e) and heat (h) as a function of $\kappa_{fb}$, for the value of $\tau_1 (=\tau_3)$ indicated by the vertical lines in the corresponding contour plots. The green and blue lines correspond to the values that maximize efficiency and work respectively. The red lines correspond to the quantities calculated considering stationary states as in Fig. \rp{Fig3}.
The plot in the third column report the values of efficiency (c), work (f) and heat (i) as a function of $\tau_1 (=\tau_3)$, for the value of $\kappa_{fb}$ indicated by the horizontal lines in the corresponding contour plots.
The other parameters are the same as those used in Fig.\ref{Fig4}.
}
\label{Fig5}
\end{figure}

\begin{figure}[t] \centering
\includegraphics[width=15.5cm]{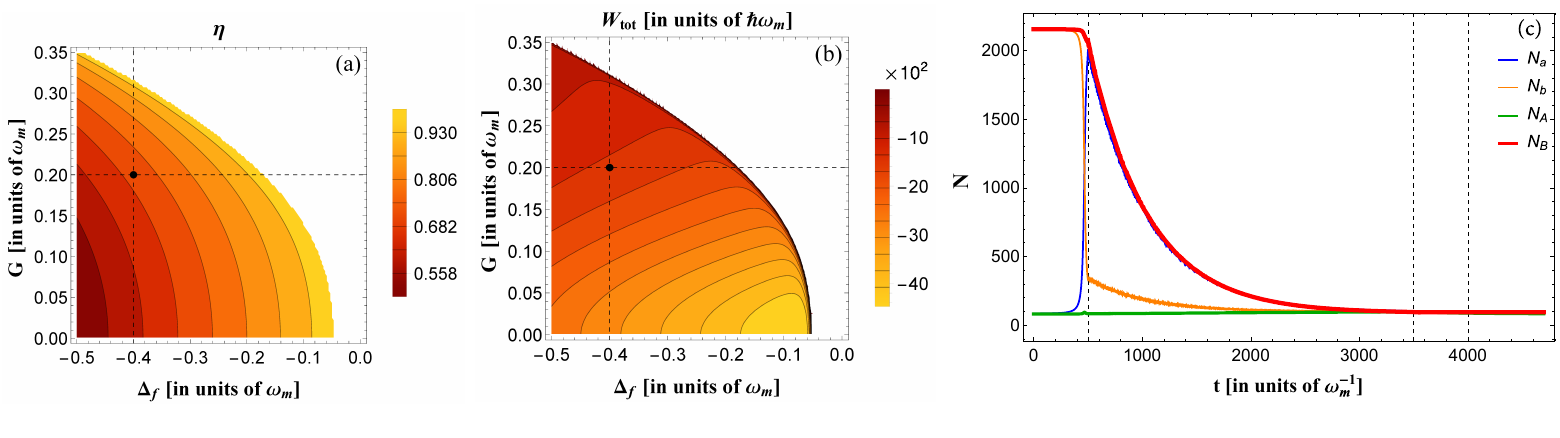}
\caption{
(a) Thermal efficiency $\eta$ and (b) total work $W_{tot}$ done by the engine as in Fig.~\ref{Fig3} but with $2 \gamma = 10^{-7}\,\omega_m$, and $n_{th} = 5000$,
$\Delta_{i} = -10\omega_m$, $ 2 \kappa_{fb}=2 \times 10^{-3}\,\omega_m$ ($n_{opt,fb}\approx 80$).
(c) the time evolution of the population of the polariton and bare modes for the parameters corresponding to the dot in plot (a) and (b) and with $\tau_1=\tau_3=500\,\omega_m^{-1}$, $\tau_2=3/\kappa_{fb}$ and $\tau_4=20/\gamma$.
}
\label{Fig6}
\end{figure}
\begin{figure}[t] \centering
\includegraphics[width=15.5cm]{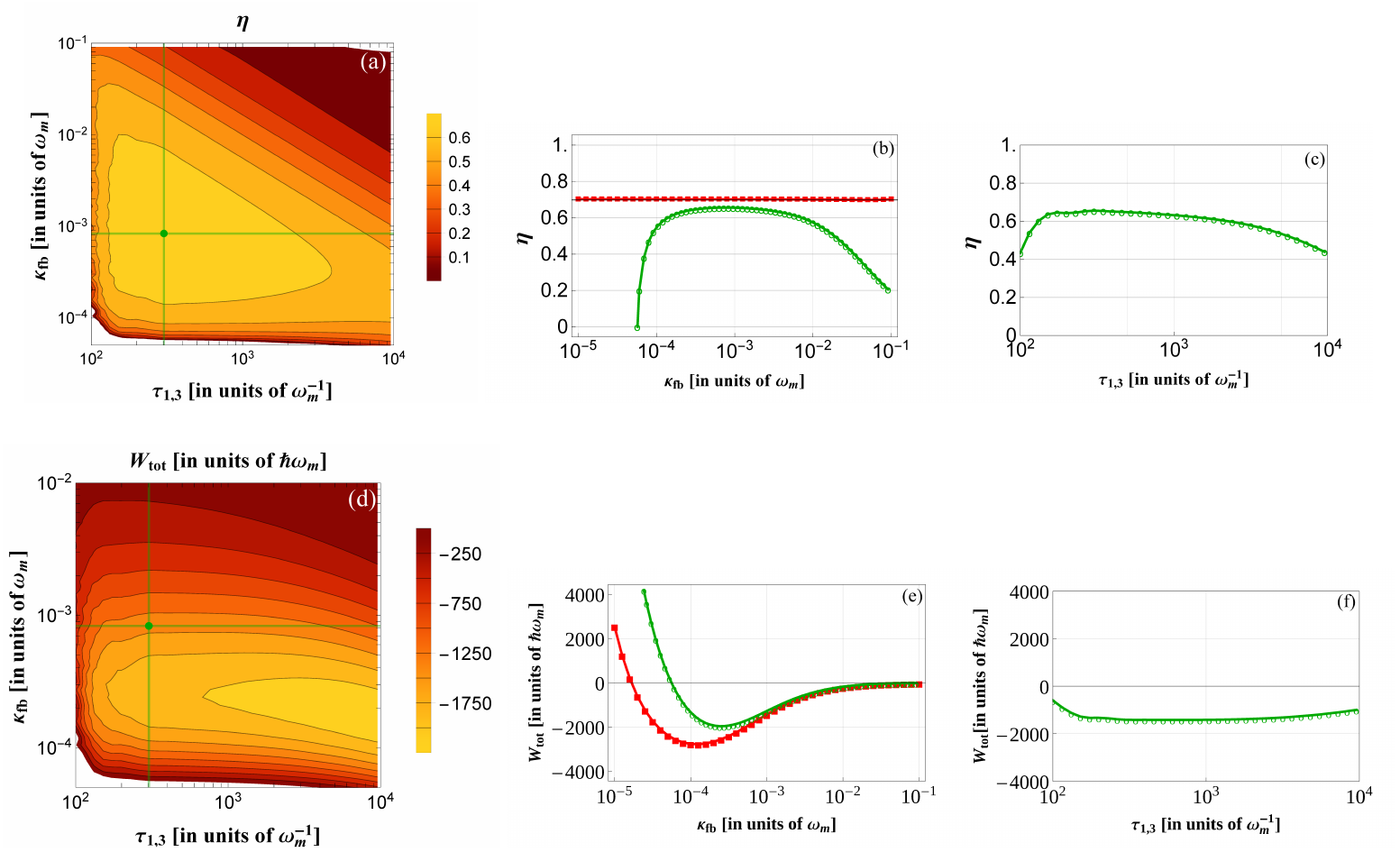}
\caption{
(a),(b),(c): Efficiency of the quantum engine; (d),(e),(f): total work versus $\kappa_{fb}$ and $\tau_1\ (=\tau_3)$, evaluated by computing the system time evolution as discussed in~\ref{heat_work}.
The plot in the second column report the values of efficiency (b) and work (e) as functions of $\kappa_{fb}$, for the value of $\tau_1 (=\tau_3)$ indicated by the vertical lines in the contour plots. The red lines correspond to the quantities calculated considering stationary states as in Fig.\rp{Fig6}.
The plot in the third column report the values of efficiency (c) and work (f) as a function of $\tau_1 (=\tau_3)$, for the value of $\kappa_{fb}$ indicated by the horizontal lines in the contour plots.
The other parameters are the same as those corresponding to Fig. \ref{Fig6}(c).
}
\label{Fig7}
\end{figure}

In Fig. \ref{Fig4} we report the results evaluated for an optomechanical coupling $G$ of the same order of the cavity decay rate $\kappa_c$. In this case the engine described in~\cite{zhang2014a,zhang2014} has low efficiency. Here we utilize feedback to effectively reduce the cavity linewidth and reach the regime of strong coupling~\cite{rossi2018} and significantly enhance the performance of the engine.
Fig. \ref{Fig4}(a) shows that the lowest polariton $B$ plays the main role in the dynamics of the system, and we can ignore the dynamics of the polariton $A$ as the variation of its excitations is relatively small during each stroke of the Otto cycle.
At the beginning of the process, the number of excitations of the upper and lower polariton modes, $N_A$ and $N_B$, are almost equal to the number of photons $N_a$ and phonon $N_b$
respectively.
The numbers of polaritons $N_A$ and $N_B$ remain almost constant during the first adiabatic passage (while at the same time the phonon and photon numbers exchange their values).  In the second stroke
the photon-like $B$-polaritons decay due to cavity dissipation (the oscillations of the photon and phonon populations are due to the optomechanical coupling). In the third stroke the polariton mode $B$ comes back to its phonon-like character, and then it slowly thermalizes to its initial population. The final thermalization is shown in plots (b) and (e) because it is very slow due to the small mechanical damping rate $\gamma$.
This behaviour is consistent with that of an Otto cycle as discussed in Section~\ref{feedbak_engine}~\cite{zhang2014a,zhang2014}.
It is also worth to notice that the population of polariton $A$ remains small throughout the whole cycle, indicating that it plays a minor role in the energy exchanges and hence in the functioning of the engine. Fig. \ref{Fig4}(c) shows the changes in energy, work and heat in each stroke.
As expected for an Otto cycle, the first and third strokes (the adiabatic passages) are mainly associated
with work production, while heat is exchanged mainly in the isochoric processes (second and fourth strokes). In particular, the system produces work in the first stroke, while it absorbs heat in the fourth stroke. The marginal imperfections of Fig. \ref{Fig4}(c) (finite heat exchanges in the first and third stroke) are due to non-ideal adiabatic processes~\cite{deangelis2018}.
As a comparison we plot in Fig.~\ref{Fig4}(d), (e) and (f) the corresponding results achievable with the same system when the feedback is off.
In this case the cavity dissipation is too large and
the population of the polariton mode $B$ decreases significantly in the first stroke, the work performed is strongly reduced and the corresponding thermal efficiency is much lower.

\begin{figure}[t] \centering
\includegraphics[width=15.5cm]{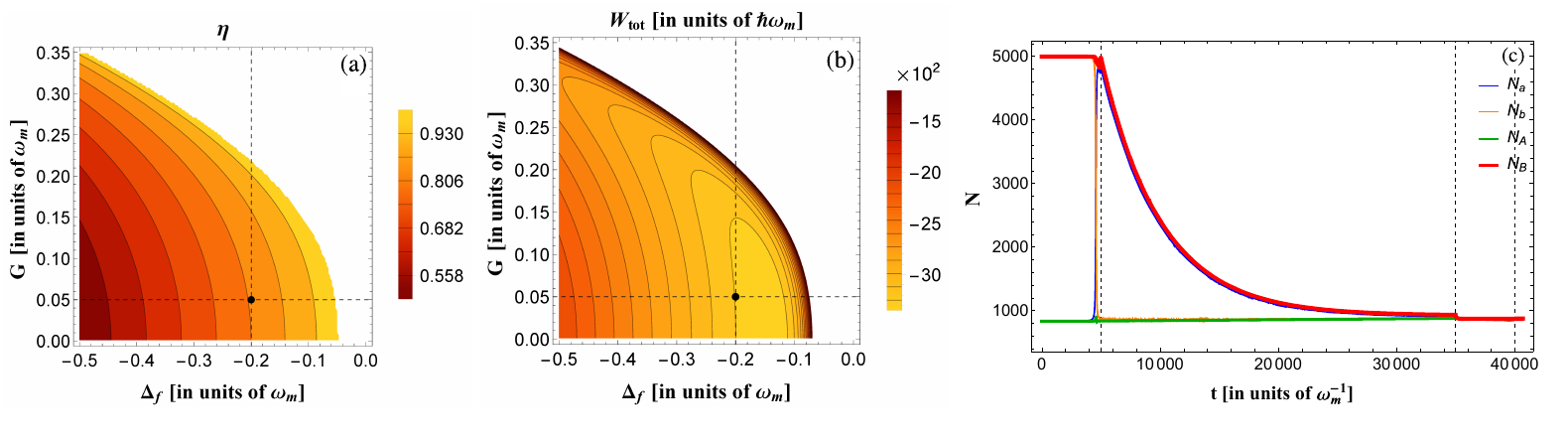}
\caption{
The same as in Fig~\ref{Fig6} with $ 2 \kappa_{fb}=2 \times 10^{-4}\,\omega_m$ ($n_{opt,fb} \approx 830$) and $\tau_1=\tau_3=5000\omega_m^{-1}$.
}
\label{Fig8}
\end{figure}
\begin{figure}[t] \centering
\includegraphics[width=15.5cm]{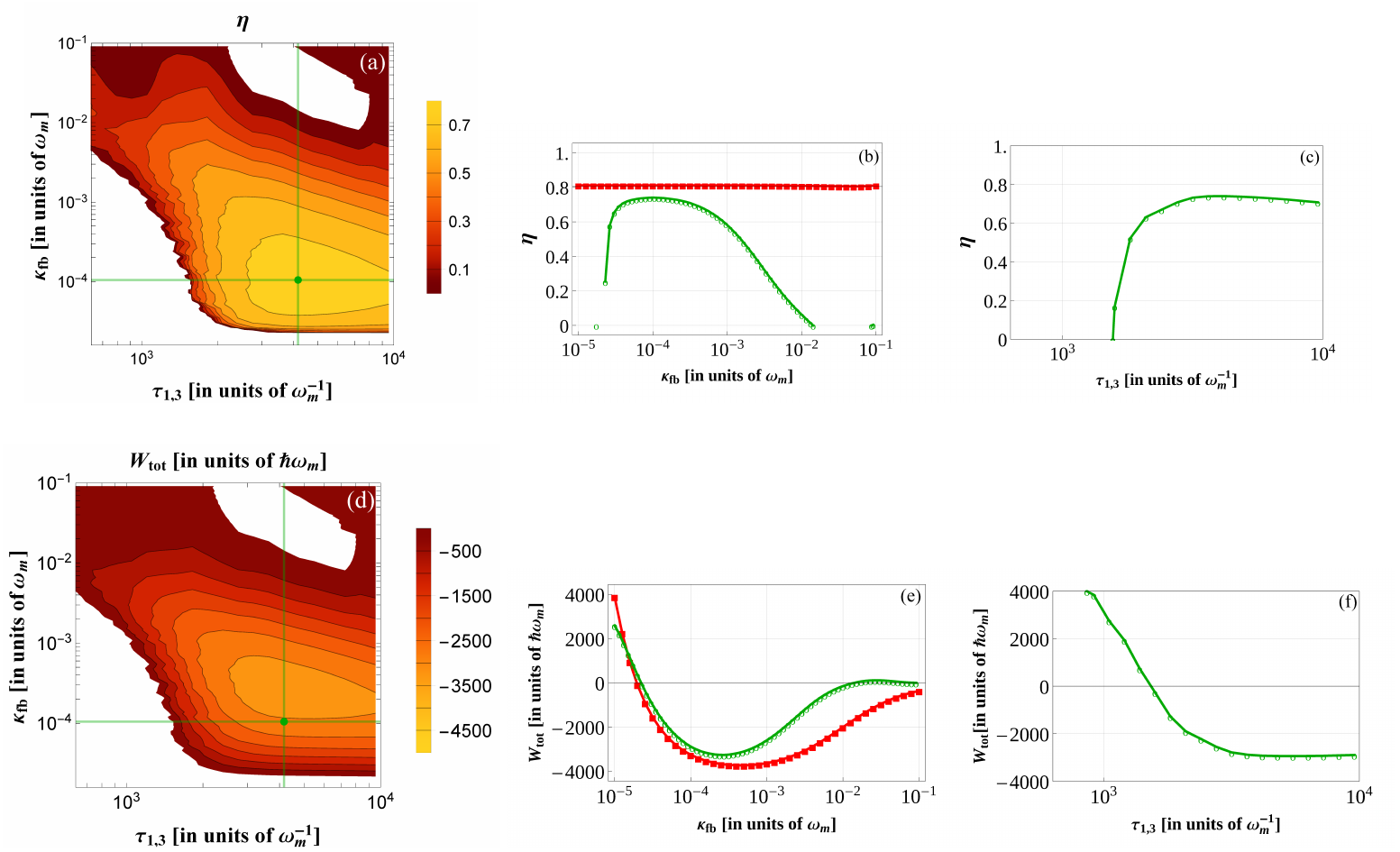}
\caption{The same as in Fig.~\ref{Fig7} but for the parameters corresponding to Fig. \ref{Fig8}(c).
}
\label{Fig9}
\end{figure}

It is instructive to analyze the performance of the engine in terms of its efficiency, performed work and absorbed heat as a function of the two most critical time scales of the engine dynamics, namely the effective decay rate $\kappa_{fb}$ and the duration of the adiabatic processes $\tau_1\ (=\tau_3)$. These results are shown in Figs.~\rp{Fig5}.
The contour plots highlight that although maximum efficiency and maximum work are not achieved for the same parameters (see the dots in the contour plots),
the corresponding values are relatively stable and the results achieved when $\eta$ is maximum are very close to those corresponding to maximum $-W_{tot}$.
The work is maximized at intermediate values of both $\kappa_{fb}$ and $\tau_1$, as a compromise between the opposite requirements described by the hierarchy relations~\rp{hi}.
The white areas in the contour plots indicate the parameters in which the engine is not functional. Namely for these parameters the total work become positive (indicating that the work is done on the system and not by the system).
Plots (b), (c), (e), (f), (h) and (i) represent the values of $\eta$, $W_{tot}$ and $Q_{abs}$, along the vertical and horizontal lines depicted in the contour plots.
The red lines in the plots (b), (e) and (h) correspond to the approximate results evaluated following the procedure used also for the results reported in Fig.~\ref{Fig3}. We observe that the exact result approaches the estimates close to the optimal values.

The work done by this engine can be easily increased by using an higher temperature phonon reservoir. This is shown in Fig.~\ref{Fig6}, where the number of thermal excitations is increased with respect to the situation of Fig.~\ref{Fig3}. In these results we have also considered a larger value of the initial detuning, which implies a longer time of the adiabatic processes, and in turn requires a smaller value of the cavity decay rate.
The corresponding time evolution of the populations of the system modes is shown in Fig.~\ref{Fig6} (c) and describes the expected behaviour discussed in section~\ref{heat_engine}.
Fig.~\ref{Fig7} instead displays the corresponding results as a function of the effective cavity decay rate $\kappa_{fb}$  and of the duration of the adiabatic processes and $\tau_1\ (=\tau_3)$ evaluated by solving the dynamics of the full model. We observe that while the efficiency of the engine is only slightly larger than the one achieved with the parameters of Fig.~\ref{Fig4},
the work done by the system is significantly larger, and it achieves its optimal value when both $\kappa_{fb}$ and $\tau_1$ fulfill the relations~\rp{hi}.

Fig.~\ref{Fig3} and \ref{Fig6} show that the optimal performance of the engine is expected for small optomechanical coupling $G$ and small final detuning $\Delta_f$, and this is confirmed by the results reported in Figs.~\ref{Fig8} and \ref{Fig9}. For the parameters used in Fig~\ref{Fig8} the system follows more closely the ideal transformations described in section~\ref{heat_engine}. Specifically Fig.~\ref{Fig8}(c) shows the time evolution of the populations of the system modes, with an almost perfect exchange of excitations between cavity and mechanical resonator in the first stroke and with $N_A$ which remains essentially constant.
Fig.~\ref{Fig9} shows that both the efficiency and the work done by the engine are significantly enhanced even if the system (without feedback) is not strongly coupled.

\section{The Otto cycle on the upper polariton}\label{upper}

\begin{figure}[t] \centering
\includegraphics[width=7cm]{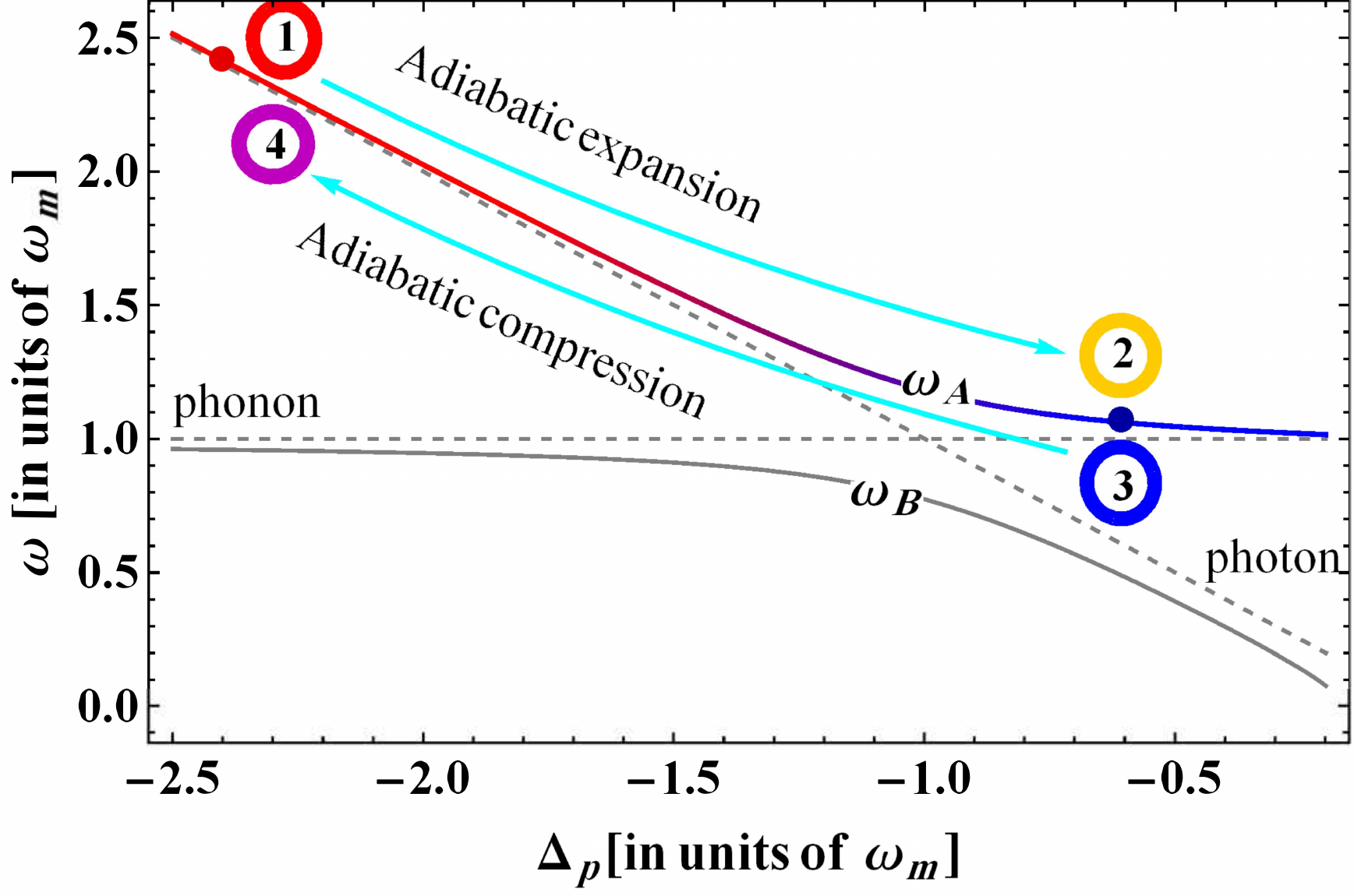}
\caption{
Scheme of the Otto cycle operated on the upper polariton (see Fig.~\ref{Fig2} for the cycle operated on the lower polariton).
}
\label{Fig10}
\end{figure}

In the previous section we have studied the thermodynamical properties of an Otto cycle operated on the lower polariton $B$.
A similar device can be implemented also using the upper polariton $A$ if
the following conditions are fulfilled
\begin{eqnarray}\label{hi2}
&&\frac{1}{G}\ll\tau_1,\tau_3 \ll \frac{1}{\gamma}\ll\tau_2\ll\frac{1}{\kappa_{fb}}\ll\tau_4\ ,\\
&&\hspace{1.3cm}n_{opt,fb} > n_{th}\ ,
\end{eqnarray}
such that the cavity effectively decay over the longest time scale 
and is coupled to the hot bath, meaning that the roles of the hot and cold bath are now exchanged.
These conditions can be, in principle, realized with the help of feedback in a system with not to small mechanical dissipation rate $\gamma$ and low thermal fluctuations. The four strokes of the cycle are then similar to what we have discussed above, and can be realized with a similar variation of the detuning~\rp{Delta}, but with the roles of photonic and phononic excitations exchanged (see Fig.~\ref{Fig10}).
Similar to our previous discussion, in this case, we can estimate an engine efficiency of $\eta\sim 1-\omega_m/\abs{\Delta_f}$.
An example of the performance of this engine is reported in Fig.~\rp{Fig11}. The results reported in Fig.~\rp{Fig11} (a) and (b) are estimates evaluated in terms of the steady state energy of the upper polariton mode at each node of the stroke (assuming the population of the lower polariton mode constant).
The efficiency in plot (a) is almost constant as a function of the final detuning $\Delta_f$ and this is due to the fact that the frequency of the upper polariton mode is almost constant for values of the detuning close to the cavity resonance (see Fig.~\ref{Fig10}).
The plots in Fig.~\rp{Fig11} (c) and (d) are the time evolution of the modes populations and the heat and work corresponding to each stroke of the cycle for the parameters indicated by the dot in plots (a) and (b), and computed using the formulas presented in \ref{heat_work}. They are qualitatively similar to the results of Fig.~\ref{Fig4} (a) and (c) and demonstrate that for the chosen parameters the system is able to transform heat into work following an Otto cycle that involves only the upper polariton.

We finally remark that we have not found any significant qualitative and quantitative difference between the two schemes in the parameter regime that we have analyzed (note that the higher efficiency in Fig.~\rp{Fig11}(d) as compared to Fig.~\rp{Fig4}(c) is due to the larger number of excitations used in the former). Furthermore we note that the scheme based on the upper polariton requires to run the feedback closer to the instability in order to achieve a sufficiently small $\kappa_{fb}$ and a sufficiently large value of $n_{opt,fb}$. This can make the experimental implementation of this engine significantly more problematic as compared to the engine based on the lower polariton.

\begin{figure} \centering
\includegraphics[width=12.5cm]{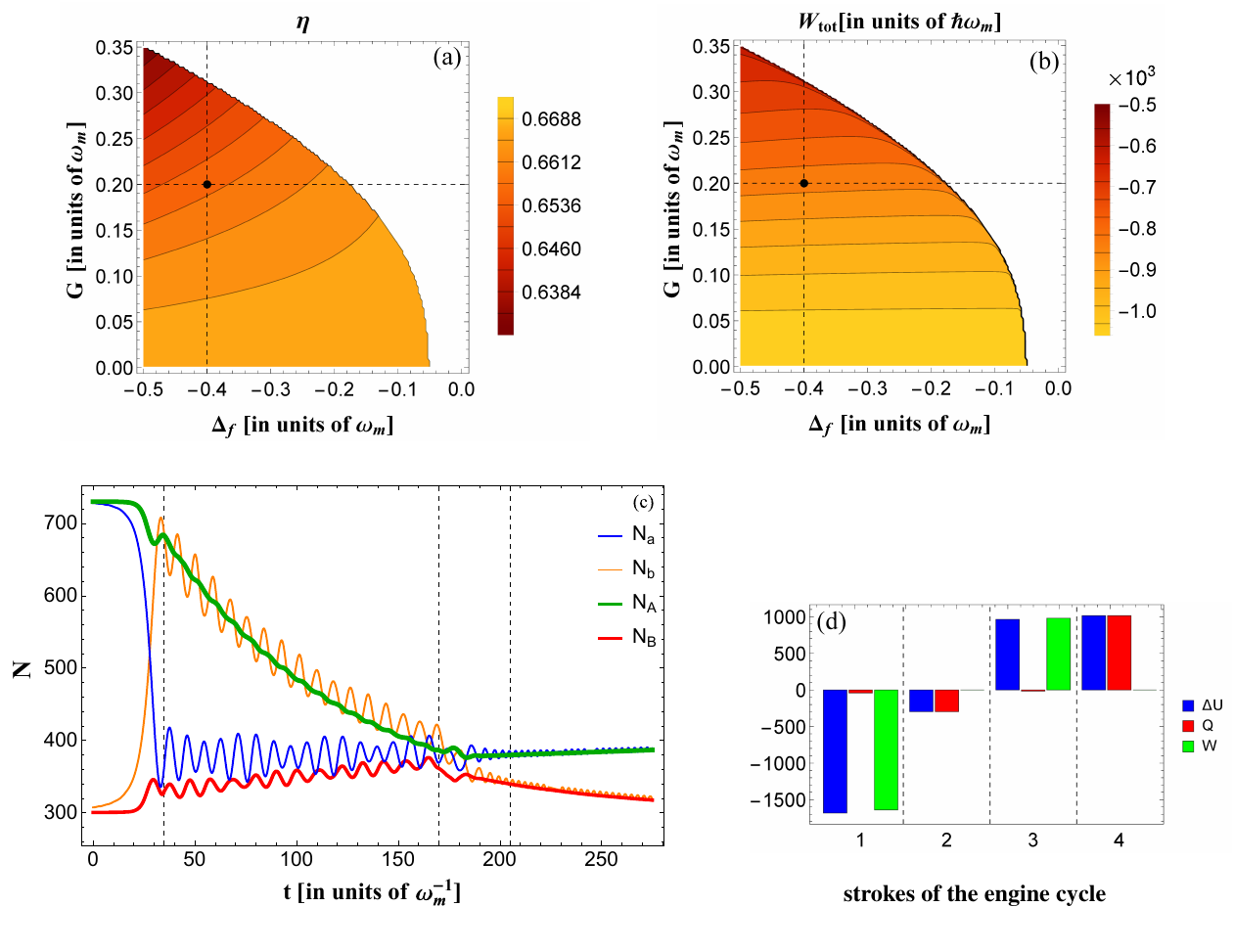}
\caption{
(a) Thermal efficiency $\eta$ and (b) total work $W_{tot}$ done by the engine
(operated on the upper polariton) as a function of the Detuning $\Delta_{f}$ and the optomechanical coupling strength $G$, evaluated in terms of the steady-state quantities corresponding to the upper polariton at each node of the cycle, and assuming perfect adiabatic processes.
The white ares indicate the parameters at which the system is unstable (see~\ref{heat_work}).
The dots indicate the parameters used for the results in plots (c) and (d). (c) Shows the
time evolution of the populations of the polariton and bare modes. (d) Shows the
corresponding energy changes (blue), heat exchanged (red) and work performed (green) during each stroke of the cycle. The duration of each stroke is $\tau_1=\tau_3 = 35 \omega_m^{-1}$,  $\tau_{2} = 135 \omega_m^{-1}$ and  $\tau_{4} = 20/\kappa_{fb} $.
The other parameters are $\Delta_{i} = -3\omega_m$, $2 \kappa_c = 0.1 \omega_m$, $2 \gamma = 0.012\, \omega_m$, and $n_{th} = 300$. The feedback is set in order to achieve the effective cavity decay rate $ 2 \kappa_{fb}=2 \times 10^{-4}\,\omega_m$ and the effective number of thermal photons $n_{opt,fb} \approx 830$.
}
\label{Fig11}
\end{figure}

\section{Conclusion}\label{conclusion}

Optomechanical devices come in very different sizes and configurations~\cite{aspelmeyer2014}. Their very high quality factor and the corresponding low natural mechanical decay rate, which is by far the lowest rate in the system dynamics, make them very versatile systems which are potential candidates for the experimental investigation of quantum thermodynamical effects. However, in spite of the many proposal of optomechanical based heat engine no experiment has demonstrated such devices so far. It is therefore important to suggest strategies for the realization of a working optomechanical quantum engine.  
Here we have shown that the experimental realization of the polariton-based quantum heat engine proposed in Refs.~\cite{zhang2014a,zhang2014,dong2015} can be significantly eased by means of a feedback system~\cite{rossi2017,kralj2017,rossi2018,zippilli2018} which allows to control the decay rate of the optical cavity.
This engine exploits the lower polariton mode as working fluid and works between the hot phononic thermal reservoir and the cold photonic reservoir with which the polariton comes into contact as the cavity pump detuning is varied around the red mechanical sideband frequency.
A critical requirement in this device is the strong coupling regime, that corresponds to an optomechanical interaction strength larger then the cavity decay rate so that the polariton modes can be resolved. 
In general the coupling strength can be controlled by tuning the driving light power. While, in principle, this could allow to achieve the strong coupling regime, in practice it is often not possible to employ the needed power due to the onset of unwanted non-linear effects. This is where the feedback realized in~\cite{rossi2018} can be helpful. 

In this work, we have reported a detailed analysis of the performance of the engine when the feedback is employed to effectively reduce the cavity decay rates by driving the system close to the feedback instability threshold.
We have demonstrated that the engine can work efficiently even if the system  without feedback is not strongly coupled (such that in absence of feedback the polariton modes are not resolved).
We have also shown that the feebdack noise, which can be seen as an effective non-zero temperature photonic bath, can be employed to define a similar engine working on the upper polariton mode where the role of the hot and cold baths are exchanged such that the feedback noise is absorbed as heat and transformed into usable work.

The feedback strategy that we have analyzed seems easily applicable in any optomechanical system since it requires optical equipment already in use in most of optomechanical experiments.
The results that we have presented correspond to systems in the resolved sideband regime and in cryogenic environments (considering a $1$MHz resonator the results in Figs.\ref{Fig3}, \ref{Fig4}, \ref{Fig5} and \ref{Fig11} would correspond to an external temperature of $100$mK, the results of Figs.~\ref{Fig6}-\ref{Fig9}, instead, would correspond to $1.7$K). Many experimental setups, both in the optical or microwave regimes, can be employed for demonstrating our proposal as for example~\cite{Groblacher2009a,Teufel2011,palomaki2013,peterson2016,clark2017,shomroni2019}. 
In order to test the efficiency of this device one should be able to measure the energy variations and to distinguish the contributions due to heat and work. This can be done by measuring the correlation matrix of the system by following for example the approach realized in~\cite{palomaki2013}.

To conclude, we highlight that although we have not discussed specific quantum effects, the system that we have studied can be used to study such phenomena. An important example is the investigation of the effects of correlations in the reservoirs which have been predicted to enhance the efficiency of a quantum heat engine beyond the Carnot limit~\cite{niedenzu2018}. This could be in principle analyzed with our optomechanical system by using, for example, a squeezed field to drive the cavity~\cite{asjad2016,clark2017}. Even more interestingly, in our case the bath correlations could be provided by the feedback loop itself~\cite{zippilli2018}. Another related and important question is whether, correlations in the working fluid as well could be employed to enhance the efficiency of the engine as discussed in~\cite{klatzow2019}. In our system, in fact, the feedback induced parametric term, which is negligible in the parameter regime that we have considered, could produce additional quantum coherence in the polariton state which may play a relevant role in certain situations. 
Finally, it is also interesting to ponder if, in some parameter regime, the behaviour of our engine could be interpreted as an instance of a Maxwell's demon~\cite{lloyd1997} which, in fact, can be seen as a feedback system.

\ack

We acknowledge the support of the European Union Horizon 2020 Programme for Research and Innovation through the Project No. 732894 (FET Proactive HOT) and the Project QuaSeRT funded by the QuantERA ERA-NET Cofund in Quantum Technologies.

\appendix

\section{The model in matrix form and the correlation matrix}\label{matrices}

The quantum Langevin equations~\rp{dotb} and ~\rp{dota} can be rewritten in matrix form, in terms of the vector of operators $\underline {\hat a^{T}}  = (\hat a, \hat b, \hat a^{\dagger}, \hat b^{\dagger})$, as
\begin{eqnarray}\label{MLE}
\underline {\dot {\hat a}}  &=& \mathcal{M} \underline{\hat a} + \mathit{Q} \underline {\hat a_{in} },
\end{eqnarray}
where the drift matrix is
\begin{eqnarray}\label{MM}
\mathcal{M} &=&  - \left( {\begin{array}{*{20}c}
   {\kappa_{fb}  - i\Delta _p } & {iG} & \kappa_c - \kappa_{fb} & {iG}  \\
   {iG} & {\gamma  + i\omega_m } & {iG} & 0  \\
 \kappa_c - \kappa_{fb} & { - iG} & {\kappa_{fb}  + i\Delta _p } & { - iG}  \\
   { - iG} & 0 & { - iG} & {\gamma  - i\omega_m }  \\
\end{array}} \right),
\end{eqnarray}
the matrix $\QQ$ is given by
\begin{eqnarray}\label{Q}
\mathit{Q} &=& \left( {\begin{array}{*{20}c}
   {\sqrt {2\kappa_{fb} } } & 0 & 0 & 0  \\
   0 & {\sqrt {2\gamma } } & 0 & 0  \\
   0 & 0 & {\sqrt {2\kappa_{fb} } } & 0  \\
   0 & 0 & 0 & {\sqrt {2\gamma } }  \\
\end{array}} \right),
\end{eqnarray}
and $\underline {\hat a_{in} }$ is the vector of noise operator $\underline {\hat a_{in}^{T}}  = (\hat a_{in}, \hat b_{in}, \hat a_{in}^{\dagger}, \hat b_{in}^{\dagger})$. From Eq.(\ref{MLE}) one finds that the evolution of the correlation matrix
\begin{eqnarray}\label{C}
\mathcal{C}  &=& \langle \underline{\hat a}\ \underline{\hat a}^{T} \rangle = \left( {\begin{array}{*{20}c}
   {\langle {\hat a\hat a}\rangle } & {\langle {\hat a \hat b} \rangle } & {\langle {\hat a\hat a^{\dagger}}\rangle } & {\langle {\hat a\hat b^{\dagger}} \rangle }  \\
   {\langle {\hat b\hat a} \rangle } & {\langle {\hat b\hat b} \rangle } & {\langle {\hat b\hat a^{\dagger}} \rangle } & {\langle {\hat b\hat b^{\dagger}} \rangle }  \\
   {\langle {\hat a^{\dagger} \hat a} \rangle } & { \langle {\hat a^{\dagger} \hat b}\rangle } & {\langle {\hat a^{\dagger} \hat a^{\dagger}} \rangle } & {\langle {\hat a^{\dagger} \hat b^{\dagger}} \rangle }  \\
   {\langle {\hat b^{\dagger} \hat a} \rangle } & {\langle {\hat b^{\dagger} \hat b} \rangle } & {\langle {\hat b^{\dagger} \hat a^{\dagger}} \rangle } & {\langle {\hat b^{\dagger} \hat b^{\dagger}} \rangle }  \\
\end{array}} \right),
\end{eqnarray}
is given by
\begin{eqnarray}\label{CC}
\dot{\mathcal{C}} &=& \mathcal{M} \mathcal{C} + \mathcal{C} \mathcal{M}^{T} + \mathit{Q} \mathcal{C}_{in}  \mathit{Q},
\end{eqnarray}
where $\mathcal{C}_{in}$ is the correlation matrix of the noise operators
\begin{eqnarray}\label{Cin}
\mathcal{C}_{in} &=& \langle \underline {\hat a_{in} }\ \underline {\hat a_{in} }^{T} \rangle =  \left( {\begin{array}{*{20}c}
   0 & 0 & {n_{opt,fb}  + 1} & 0  \\
   0 & 0 & 0 & {n_{th}  + 1}  \\
   {n_{opt,fb} } & 0 & 0 & 0  \\
   0 & {n_{th} } & 0 & 0  \\
\end{array}} \right)
\end{eqnarray}
(note that in the absence of the feedback we have $\kappa_{fb} = \kappa_c$ and $n_{opt,fb}=0$).

By defining $\mathcal{N} = \mathit{Q} \mathcal{C}_{in} \mathit{Q}$ and introducing the linear super operator $\hat{\mathcal{L}}$ so that $\hat{\mathcal{L}} \mathit{C} = \mathcal{M} \mathit{C}+\mathit{C} \mathcal{M}^{T}$ we can find the stationary correlation matrix
\begin{eqnarray}\label{Css}
\mathcal{C}_{ss} &=& - \hat {\mathcal{L}}^{-1} \mathcal{N}\ .
\end{eqnarray}

\section{Polariton description of the system}\label{polariton}

The Hamiltonian corresponding to the quantum Langevin equations.~\rp{dotb} and \rp{dota} is
\begin{eqnarray}\label{Hfb}
\hat H_{fb}=  - \hbar \Delta _p\ \hat a^{\dagger} \hat a + \hbar \omega _m \hat b^{\dagger} \hat b + \hbar G(\hat b + \hat b^{\dagger} )(\hat a + \hat a^{\dagger})-\ii\,\hbar\,\frac{\kappa_c-\kappa_{fb}}{2}\pp{\hat a^2-\hat a\da{}^2}\ .
\end{eqnarray}
The uncoupled normal modes of $\hat H_{fb}$, i.e. the polariton modes,
can be expressed in terms of the bare operators $\hat a $ ($\hat a^{\dagger} $) and  $\hat b $ ($\hat b^{\dagger} $) through a transformation matrix $\mathcal{T}$ as
\begin{eqnarray}\label{plo1}
\underline {\hat A} &=& \mathcal{T}^{ - 1} \underline {\hat a},
\end{eqnarray}
where $\underline {\hat A}  = (\hat A,\hat B,\hat A^{\dagger} ,\hat B^{\dagger})^T $ is the vector of the polariton operators
and $\mathcal{T}$ is a symplectic transformation that satisfies the relations
 $\mathcal{T} \mathcal{I} \mathcal{T}^{T}=\mathcal{I}$ and 
$\GG\ \TT\ \GG=\TT^*$ 
where
$\TT^*$ is the matrix with the complex conjugate elements of $\TT$, $\GG= \left( {\begin{array}{*{20}c}
   0 & \mathbf{1}  \\
   { \mathbf{1}} & 0  \\
\end{array}} \right)$,
and $\mathcal{I}$ is the symplectic form
$\mathcal{I} = \left( {\begin{array}{*{20}c}
   0 & \mathbf{1}  \\
   { - \mathbf{1}} & 0  \\
\end{array}} \right)$,
with
$\mathbf{1}$ the identity matrix.
In terms of the polariton operators, the Hamiltonian~\rp{Hfb} reads
\begin{eqnarray}\label{H0}
\hat H_{fb} &=& \hbar \omega _A \hat A^{\dagger} \hat A + \hbar \omega _A \hat B^{\dagger} \hat B + const ,
\end{eqnarray}
and the transformation matrix $\mathcal{T}$ can be obtained by solving the eigenvalue problem
\begin{eqnarray}\label{eig1}
\mathcal{M}_0 \: \mathcal{T} &=& \mathcal{T} \mathcal{D},
\end{eqnarray}
where, $\mathcal{M}_0 = \mathcal{I} \mathcal{H}$ with $ \mathcal{H} $ the matrix representation of Eq.(\ref{Hfb}), i.e., $\hat H_{fb} = \underline {\hat a}^{T} \mathcal{H} \underline {\hat a} $, given by
\begin{eqnarray}\label{HM}
\mathcal{H} &=& \frac{\hbar }{2}\left( {\begin{array}{*{20}c}
   - \ii\pp{\kappa_c-\kappa_{fb}} & G & { - \Delta _p } & G  \\
   G & 0 & G & {\omega_m }  \\
   { - \Delta _p } & G & \ii\pp{\kappa_c-\kappa_{fb}} & G  \\
   G & {\omega_m } & G & 0  \\
\end{array}} \right),
\end{eqnarray}
and $\mathcal{D}$ the diagonal matrix of symplectic eigenvalues, defined as $\mathcal{D}=\frac{1}{2}\ {\rm diag}\pg{\omega_A,\omega_B,-\omega_A,-\omega_B}$ where
\begin{eqnarray}\label{energy}
 \omega _A  &=& \frac{1}{{\sqrt 2 }}\sqrt {\Delta _p^2 -\pp{\kappa_c-\kappa_{fb}}^2 + \omega_m^2 + \sqrt {\pq{ \Delta _p^2 -\pp{\kappa_c-\kappa_{fb}}^2 - \omega_m^2 }^2  - 16 G^2 \Delta _p \omega_m } } \: , \\
 \omega _B  &=& \frac{1}{{\sqrt 2 }}\sqrt {\Delta _p^2 -\pp{\kappa_c-\kappa_{fb}}^2 + \omega_m^2 - \sqrt {\pq{ \Delta _p^2 -\pp{\kappa_c-\kappa_{fb}}^2 - \omega_m^2 }^2  - 16 G^2 \Delta _p \omega_m } } \: .
\end{eqnarray}
In Fig.\ref{Fig2}, we have depicted these eigenfrequencies in the red detuning regime ($\Delta_p < 0$) where the beam-splitter interaction term of the Hamiltonian of Eq.(\ref{Hfb}) plays the dominant role~\cite{zhang2014a}.
The Hamiltonian~\rp{Hfb} is stable, and the polariton modes can be defined, whenever the lowest eigenfrequency $\omega_B$ is real positive, i.e. when
$\Delta_p < -2 G^2/\omega_m-\sqrt{4 G^4/\omega_m^2+\pp{\kappa_c-\kappa_{fb}}^2}$.

The correlation matrix $\mathcal{C}_p$ for the polariton modes
\begin{eqnarray}\label{Cp}
\mathcal{C}_p  &=& \langle \underline{\hat A}\ \underline{\hat A}^{T} \rangle = \left( {\begin{array}{*{20}c}
   {\langle {\hat A\hat A}\rangle } & {\langle {\hat A\hat B} \rangle } & {\langle {\hat A\hat A^{\dagger}}\rangle } & {\langle {\hat A\hat B^{\dagger}} \rangle }  \\
   {\langle {\hat B\hat A} \rangle } & {\langle {\hat B\hat B} \rangle } & {\langle {\hat B\hat A^{\dagger}} \rangle } & {\langle {\hat B\hat B^{\dagger}} \rangle }  \\
   {\langle {\hat A^{\dagger} \hat A} \rangle } & { \langle {\hat A^{\dagger} \hat B}\rangle } & {\langle {\hat A^{\dagger} \hat A^{\dagger}} \rangle } & {\langle {\hat A^{\dagger} \hat B^{\dagger}} \rangle }  \\
   {\langle {\hat B^{\dagger} \hat A} \rangle } & {\langle {\hat B^{\dagger} \hat B} \rangle } & {\langle {\hat B^{\dagger} \hat A^{\dagger}} \rangle } & {\langle {\hat B^{\dagger} \hat B^{\dagger}} \rangle }  \\
\end{array}} \right),
\end{eqnarray}
is related to the bare modes correlation matrix by the relation
\begin{eqnarray}\label{CpC}
\mathcal{C}_p  &=& \mathcal{T}^{-1} \mathcal{C} (\mathcal{T}^{-1})^{T}.
\end{eqnarray}
In particular the steady state in the polariton base is obtained by computing Eq.~\rp{CpC} on the steady state correlation matrix~\rp{Css}. The steady state population of the polariton $B$ is then given by the element $(4,2)$ of the resulting matrix [see~\rp{Cp}], i.e. $N_B=\pg{\CC_p}_{4,2}$, and similarly $N_A=\pg{\CC_p}_{3,1}$.

\section{Heat and work}\label{heat_work}

The internal energy $U$ of the system can be expressed in terms of the average of the system Hamiltonian~\cite{vinjanampathy2016}
\begin{eqnarray}\label{U}
U(t) &=& \langle \hat{H}_{fb}(t)\rangle =Tr[\hat \rho (t)\hat H_{fb}(t)]
\end{eqnarray}
where $\rho(t)$ is the density matrix which describes the state of the system at time $t$, and $H_{fb}(t)$ is the system Hamiltonian~\rp{Hfb},
with time dependent detuning $\Delta_p(t)$.
The energy change is given by the temporal derivative of the internal energy
\begin{eqnarray}\label{Ut}
\dot U(t) &=&  Tr\pq{\hat \rho (t) \dot{\hat H}(t)}+Tr\pq{\dot{\hat \rho} (t)\hat H(t)},
\end{eqnarray}
which is the sum of two contributions.
The first, associated with the variation of the system Hamiltonian, contributes to the work, while the second one is due to irreversible dissipative processes and contributes to the heat~\cite{alicki1979}. Specifically, in a process that takes place from the initial time $t_i$ to the final time $t_f$, the
heat $Q$ and the work $W$ are defined by the time integrals
\begin{eqnarray}\label{QWt}
Q &=& \int_{t_i}^{t_f} \dd t\ {Tr[\dot {\hat \rho} (t)\hat H_{fb}(t)]},\\
W &=& \int_{t_i}^{t_f}\dd t\ {Tr[\hat \rho (t) \dot {\hat H}_{fb}(t)]},
\end{eqnarray}
such that
\begin{eqnarray}\label{1st}
Q+ W = \Delta U
\end{eqnarray}
which represent the first law of thermodynamics.
The difference in internal energy $\Delta U$ can be computed in terms of the average values of the system Hamiltonian~\rp{Hfb}, as
 $\Delta U=\av{H_{fb}(t_f)}-\av{H_{fb}(t_i)}$. In particular, the average value $\av{H_{fb}(t)}$
can be expressed in terms of the second order correlation functions [which in turn are derived using the solution of the equation for the correlation matrix~\rp{CC} evaluated using the time dependent detuning~\rp{Delta}] as
\begin{eqnarray}\label{Hmin}
\langle \hat H_{fb}(t) \rangle &=& - \hbar \Delta _p(t) \langle {\hat a^{\dagger}(t)\ \hat{a}(t)} \rangle + \hbar \omega_m \langle {\hat b^{\dagger}(t)\ \hat{b}(t)} \rangle
\nn\\&&
+ \hbar G \pq{\langle {\hat b(t)\ \hat{a}(t)} \rangle + \langle {\hat b(t)\ \hat{a}^{\dagger}(t)} \rangle + \langle \hat b^{\dagger}(t)\ \hat{a}(t) \rangle + \langle \hat b^{\dagger}(t)\ \hat{a}^{\dagger}(t) \rangle }
\nn\\&&
-\ii\,\hbar\,\frac{\kappa_c-\kappa_{fb}}{2}\pq{\av{\hat a(t)^2}-\av{\hat a\da(t)^2}}\ .
\end{eqnarray}
The heat, instead, can be computed by substituting
the system Hamiltonian~\rp{Hfb} and the system master equations
\begin{eqnarray}\label{master}
\dot {\hat \rho} &=& -\frac{{i}}{\hbar }[\hat H_{fb}(t),\hat \rho] + \kappa _c (n_{opt,fb}  + 1)(2\hat a\hat \rho \hat a^{\dagger}  - \hat a^{\dagger} \hat a\hat \rho  - \hat \rho \hat a^{\dagger} \hat a) + \kappa _c\ n_{opt,fb} (2\hat a^{\dagger} \hat \rho \hat a - \hat a\hat a^{\dagger} \hat \rho  - \hat \rho \hat a\hat a^{\dagger}) \nn \\
&+& \gamma (n_{th}  + 1)(2\hat b\hat \rho \hat b^{\dagger}  - \hat b^{\dagger} \hat b\hat \rho  - \hat \rho \hat b^{\dagger} \hat b) + \gamma n_{th} (2\hat b^{\dagger} \hat \rho \hat b - \hat b\hat b^{\dagger} \hat \rho  - \hat \rho \hat b\hat b^{\dagger})
\end{eqnarray}
[which provides a description of the system dynamics equivalent to the quantum Langevin equations~\rp{dotb} and \rp{dota}], into Eq.~\rp{QWt}. Thereby,
exploiting the cyclic property of the trace, one finds that the heat exchanged with the environment in a process from time $t_i$ to $t_f$ is given by
\begin{eqnarray}\label{dQ}
Q=\int_{t_i}^{t_f} \dd t\ {Tr[\dot {\hat \rho} (t)\hat H_{fb}(t)]}&=&
\int_{t_i}^{t_f} \dd t\ \lpg{
2\hbar \omega_m \gamma n_{th}  - 2\hbar \Delta _p(t) \kappa _c n_{opt}  + 2 \hbar \Delta _p(t) \kappa _c \langle {\hat a^{\dagger}(t)\ \hat{a(t)}} \rangle
}\nn\\&&
- 2 \hbar \omega_m \gamma \kappa _c \langle {\hat b^{\dagger}(t)\ \hat{b}(t)} \rangle
\nn\\&&
- \hbar G (\kappa_c + \gamma) \pq{
\langle {\hat b(t)\ \hat{a}(t)} \rangle + \langle {\hat b(t)\ \hat{a}^{\dagger}(t)} \rangle + \langle \hat b^{\dagger}(t)\ \hat{a}(t) \rangle + \langle \hat b^{\dagger}(t)\ \hat{a}^{\dagger}(t) \rangle }
\nn\\&&\rpg{
+\ii\hbar\,\kappa_c\,\pp{\kappa_c-\kappa_{fb}}\pq{\av{\hat a(t)^2}-\av{\hat a\da(t)^2}}
}\ .
\end{eqnarray}
The correlation functions in this expression can be computed by solving the equation for the correlation matrix~\rp{CC} [with the time dependent detuning].
Finally, the work is determined, in terms of these results for $\Delta U$ and $Q$, using the first law of thermodynamics~\rp{1st}.

We notice that this approach allows to extend the numerical analysis introduced in~\cite{zhang2014a,zhang2014,dong2015} (which, being based on the numerical integration of the master equation, is constrained to a low number of system excitations) to an arbitrary number of excitations.

\section*{References}

\bibliographystyle{unsrt}
\bibliography{OM_Engine_Feedback}

\begin{thebibliography}{10}

\bibitem{vinjanampathy2016}
Sai Vinjanampathy and Janet Anders.
\newblock Quantum thermodynamics.
\newblock {\em Contemporary Physics}, 57(4):545--579, October 2016.

\bibitem{alicki2018}
Robert Alicki and Ronnie Kosloff.
\newblock Introduction to {{Quantum Thermodynamics}}: {{History}} and
  {{Prospects}}.
\newblock {\em arXiv:1801.08314 [quant-ph]}, January 2018.

\bibitem{bowen2015}
Warwick~P. Bowen and Gerard~J. Milburn.
\newblock {\em Quantum {{Optomechanics}}}.
\newblock {Taylor \& Francis}, November 2015.

\bibitem{Li2018}
Bei-Bei Li, Jan B{\'i}lek, Ulrich~B. Hoff, Lars~S. Madsen, Stefan Forstner,
  Varun Prakash, Clemens Sch{\"a}fermeier, Tobias Gehring, Warwick~P. Bowen,
  and Ulrik~L. Andersen.
\newblock Quantum enhanced optomechanical magnetometry.
\newblock {\em Optica}, 5(7):850, July 2018.

\bibitem{aspelmeyer2014}
Markus Aspelmeyer, Tobias~J. Kippenberg, and Florian Marquardt.
\newblock Cavity optomechanics.
\newblock {\em Reviews of Modern Physics}, 86(4):1391--1452, December 2014.

\bibitem{Bawaj2015}
Mateusz Bawaj, Ciro Biancofiore, Michele Bonaldi, Federica Bonfigli, Antonio
  Borrielli, Giovanni Di~Giuseppe, Lorenzo Marconi, Francesco Marino, Riccardo
  Natali, Antonio Pontin, Giovanni~A. Prodi, Enrico Serra, David Vitali, and
  Francesco Marin.
\newblock Probing deformed commutators with macroscopic harmonic oscillators.
\newblock {\em Nature Communications}, 6:7503, June 2015.

\bibitem{zhang2014a}
Keye Zhang, Francesco Bariani, and Pierre Meystre.
\newblock Quantum {{Optomechanical Heat Engine}}.
\newblock {\em Physical Review Letters}, 112(15):150602, April 2014.

\bibitem{zhang2014}
Keye Zhang, Francesco Bariani, and Pierre Meystre.
\newblock Theory of an optomechanical quantum heat engine.
\newblock {\em Physical Review A}, 90(2):023819, August 2014.

\bibitem{dong2015}
Ying Dong, Keye Zhang, Francesco Bariani, and Pierre Meystre.
\newblock Work measurement in an optomechanical quantum heat engine.
\newblock {\em Physical Review A}, 92(3):033854, September 2015.

\bibitem{dong2015a}
Ying Dong, F.~Bariani, and P.~Meystre.
\newblock Phonon {{Cooling}} by an {{Optomechanical Heat Pump}}.
\newblock {\em Physical Review Letters}, 115(22):223602, November 2015.

\bibitem{dechant2015a}
Andreas Dechant, Nikolai Kiesel, and Eric Lutz.
\newblock All-{{Optical Nanomechanical Heat Engine}}.
\newblock {\em Physical Review Letters}, 114(18):183602, May 2015.

\bibitem{mari2015}
A.~Mari, A.~Farace, and V.~Giovannetti.
\newblock Quantum optomechanical piston engines powered by heat.
\newblock {\em Journal of Physics B: Atomic, Molecular and Optical Physics},
  48(17):175501, 2015.

\bibitem{gelbwaser-klimovsky2015a}
D.~{Gelbwaser-Klimovsky} and G.~Kurizki.
\newblock Work extraction from heat-powered quantized optomechanical setups.
\newblock {\em Scientific Reports}, 5:07809, January 2015.

\bibitem{bathaee2016}
M.~Bathaee and A.~R. Bahrampour.
\newblock Optimal control of the power adiabatic stroke of an optomechanical
  heat engine.
\newblock {\em Physical Review E}, 94(2):022141, August 2016.

\bibitem{zhang2017}
Keye Zhang and Weiping Zhang.
\newblock Quantum optomechanical straight-twin engine.
\newblock {\em Physical Review A}, 95(5):053870, May 2017.

\bibitem{rossi2018}
Massimiliano Rossi, Nenad Kralj, Stefano Zippilli, Riccardo Natali, Antonio
  Borrielli, Gregory Pandraud, Enrico Serra, Giovanni Di~Giuseppe, and David
  Vitali.
\newblock Normal-{{Mode Splitting}} in a {{Weakly Coupled Optomechanical
  System}}.
\newblock {\em Physical Review Letters}, 120(7):073601, February 2018.

\bibitem{rossi2017}
Massimiliano Rossi, Nenad Kralj, Stefano Zippilli, Riccardo Natali, Antonio
  Borrielli, Gregory Pandraud, Enrico Serra, Giovanni Di~Giuseppe, and David
  Vitali.
\newblock Enhancing {{Sideband Cooling}} by {{Feedback}}-{{Controlled Light}}.
\newblock {\em Physical Review Letters}, 119(12):123603, September 2017.

\bibitem{kralj2017}
Nenad Kralj, Massimiliano Rossi, Stefano Zippilli, Riccardo Natali, Antonio
  Borrielli, Gregory Pandraud, Enrico Serra, Giovanni~Di Giuseppe, and David
  Vitali.
\newblock Enhancement of three-mode optomechanical interaction by
  feedback-controlled light.
\newblock {\em Quantum Science and Technology}, 2(3):034014, 2017.

\bibitem{zippilli2018}
Stefano Zippilli, Nenad Kralj, Massimiliano Rossi, Giovanni Di~Giuseppe, and
  David Vitali.
\newblock Cavity optomechanics with feedback-controlled in-loop light.
\newblock {\em Physical Review A}, 98(2):023828, August 2018.

\bibitem{deangelis2018}
Giulia~Vittoria De~Angelis.
\newblock {\em Optomechanical Heat Engine with Feedback-Controlled Light}.
\newblock Master {{Thesis}}, Universit{\`a} degli studi di Camerino,
  {Camerino}, 2018.

\bibitem{gardiner2004}
Crispin Gardiner and Peter Zoller.
\newblock {\em Quantum {{Noise}}: {{A Handbook}} of {{Markovian}} and
  {{Non}}-{{Markovian Quantum Stochastic Methods}} with {{Applications}} to
  {{Quantum Optics}}}.
\newblock Springer {{Series}} in {{Synergetics}}. {Springer-Verlag}, {Berlin
  Heidelberg}, 3 edition, 2004.

\bibitem{Groblacher2009a}
Simon Gr{\"o}blacher, Jared~B. Hertzberg, Michael~R. Vanner, Garrett~D. Cole,
  Sylvain Gigan, K.~C. Schwab, and Markus Aspelmeyer.
\newblock Demonstration of an ultracold micro-optomechanical oscillator in a
  cryogenic cavity.
\newblock {\em Nature Physics}, 5(7):485--488, July 2009.

\bibitem{Teufel2011}
J.~D. Teufel, Dale Li, M.~S. Allman, K.~Cicak, A.~J. Sirois, J.~D. Whittaker,
  and R.~W. Simmonds.
\newblock Circuit cavity electromechanics in the strong-coupling regime.
\newblock {\em Nature}, 471(7337):204--208, March 2011.

\bibitem{palomaki2013}
T.~A. Palomaki, J.~D. Teufel, R.~W. Simmonds, and K.~W. Lehnert.
\newblock Entangling {{Mechanical Motion}} with {{Microwave Fields}}.
\newblock {\em Science}, 342(6159):710--713, August 2013.

\bibitem{peterson2016}
R.~W. Peterson, T.~P. Purdy, N.~S. Kampel, R.~W. Andrews, P.-L. Yu, K.~W.
  Lehnert, and C.~A. Regal.
\newblock Laser {{Cooling}} of a {{Micromechanical Membrane}} to the {{Quantum
  Backaction Limit}}.
\newblock {\em Physical Review Letters}, 116(6):063601, February 2016.

\bibitem{clark2017}
Jeremy~B. Clark, Florent Lecocq, Raymond~W. Simmonds, Jos{\'e} Aumentado, and
  John~D. Teufel.
\newblock Sideband cooling beyond the quantum backaction limit with squeezed
  light.
\newblock {\em Nature}, 541(7636):191--195, January 2017.

\bibitem{shomroni2019}
Itay Shomroni, Liu Qiu, Daniel Malz, Andreas Nunnenkamp, and Tobias~J.
  Kippenberg.
\newblock Optical backaction-evading measurement of a mechanical oscillator.
\newblock {\em Nature Communications}, 10(1):2086, December 2019.

\bibitem{niedenzu2018}
Wolfgang Niedenzu, Victor Mukherjee, Arnab Ghosh, Abraham~G. Kofman, and
  Gershon Kurizki.
\newblock Quantum engine efficiency bound beyond the second law of
  thermodynamics.
\newblock {\em Nature Communications}, 9(1):165, January 2018.

\bibitem{asjad2016}
Muhammad Asjad, Stefano Zippilli, and David Vitali.
\newblock Suppression of {{Stokes}} scattering and improved optomechanical
  cooling with squeezed light.
\newblock {\em Physical Review A}, 94(5):051801, November 2016.

\bibitem{klatzow2019}
James Klatzow, Jonas~N. Becker, Patrick~M. Ledingham, Christian Weinzetl,
  Krzysztof~T. Kaczmarek, Dylan~J. Saunders, Joshua Nunn, Ian~A. Walmsley, Raam
  Uzdin, and Eilon Poem.
\newblock Experimental {{Demonstration}} of {{Quantum Effects}} in the
  {{Operation}} of {{Microscopic Heat Engines}}.
\newblock {\em Physical Review Letters}, 122(11):110601, March 2019.

\bibitem{lloyd1997}
Seth Lloyd.
\newblock Quantum-mechanical {{Maxwell}}'s demon.
\newblock {\em Physical Review A}, 56(5):3374--3382, November 1997.

\bibitem{alicki1979}
R~Alicki.
\newblock The quantum open system as a model of the heat engine.
\newblock {\em Journal of Physics A: Mathematical and General},
  12(5):L103--L107, May 1979.

\end{thebibliography}

\end{document}